# Highlights

**Insights into experimental evaluation of the non-fourier heat transfer model in biological tissues**

Mohammad Azhdari, Ghader Rezazadeh, Raghav Pathak, Hans-Michael Tautenhahn, Franziska Tautenhahn, Tim Ricken, Seyed Morteza Seyedpour

- Phase lag coefficients were experimentally extracted for bovine skin, which shares thermal properties closely resembling those of human skin.

- The individual influence of each phase lag term on transient thermal responses and temperature-depth profiles was systematically investigated.

- A decoupled parameter analysis approach was employed to isolate the independent effects of each phase lag term, minimizing confounding influences from other thermal parameters.

- Results revealed that the extracted phase lag values were significantly smaller than those reported in earlier studies, highlighting the importance of precise experimental validation.

# Insights into experimental evaluation of the non-fourier heat transfer model in biological tissues

Mohammad Azhdari[a], Ghader Rezazadeh[b,c,*], Raghav Pathak[a,d], Hans-Michael Tautenhahn[e], Franziska Tautenhahn[f], Tim Ricken[a,d], Seyed Morteza Seyedpour[a,d,*]

[a]*Institute of Structural Mechanics and Dynamics in Aerospace Engineering, Faculty of Aerospace Engineering and Geodesy, University of Stuttgart, Pfaffenwaldring 27, Stuttgart, 70569, Germany*
[b]*Center for Materials Technologies, Skolkovo Institute of Science and Technology, Moscow, Russia*
[c]*Mechanical Engineering Department, Faculty of Engineering, Urmia University , 11km Sero Road, Urmia, 16557153, Iran*
[d]*Porous Media Lab, Institute of Structural Mechanics and Dynamics in Aerospace Engineering, Faculty of Aerospace Engineering and Geodesy, University of Stuttgart , Pfaffenwaldring 27, Stuttgart, 70569, Germany*
[e]*Department of Visceral, Transplantation, Thoracic and Vascular Surgery, Leipzig University Hospital, Liebigstraße 20, Leipzig, 04103, Germany*
[f] *Department of Oral, Maxillofacial and Plastic Facial Surgery, University Hospital Leipzig, Liebigstr. 12, Haus 1 , Leipzig, 04103, Germany*

**Abstract**

A comprehensive understanding of heat transfer mechanisms in biological tissues is essential for the advancement of thermal therapeutic techniques and the development of accurate bioheat transfer models. Conventional models often fail to capture the inherently complex thermal behavior of biological media, necessitating more sophisticated approaches for experimental validation and parameter extraction. In this study, the Two-Dimensional Three-Phase Lag (TPL) heat transfer model, implemented via the finite difference method (FDM), was employed to extract key phase lag parameters characterizing heat conduction in bovine skin tissue. Experimental measurements were obtained using a 450 nm laser source and two non-contact infrared sensors. The influence of four critical parameters was systematically investigated: heat flux phase lag ($\tau_q$), temperature gradient phase lag ($\tau_\theta$), thermal displacement coefficient ($k^*$), and thermal displacement phase lag ($\tau_v$). A carefully designed experimental protocol was used to assess each parameter independently. The results revealed that the extracted phase lag values were substantially lower than those previously reported in the literature. This highlights the importance of high-precision measurements and the need to isolate each parameter during analysis. These findings contribute to the refinement of bioheat transfer models and hold potential for improving the efficacy and safety of clinical thermal therapies.

*Corresponding author

## 1. Introduction

The importance of Non-Fourier modeling in bioheat transfer becomes particularly evident in scenarios that involve rapid and dynamic temperature changes within biological tissues, such as those encountered during thermal therapeutic procedures. Classical Fourier's law of heat conduction presumes instantaneous heat transfer, implying an infinite propagation speed of thermal waves. However, this assumption does not hold true for biological tissues, which are complex and inherently heterogeneous in nature [1, 2]. Biological systems encompass diverse structural and physiological features, including intricate cellular matrices, complex vascular networks, interstitial fluid matrices, blood perfusion mechanisms, and continuous metabolic heat generation. Each of these elements contributes toplexity and induces significant delays and inertia in the thermal response of biological tissues, aspects that Fourier theory does not adequately capture [3–7]. Numerous experimental and theoretical investigations have consistently shown that Fourier's law is insufficient for predicting the thermal responses of biological tissues subjected to rapid heating or cooling scenarios. This inadequacy becomes particularly apparent in clinical and therapeutic contexts, such as short-duration laser irradiations and cryosurgery, where the instantaneous energy transfer implied by Fourier theory significantly contradicts observed delayed thermal responses [7–11]. To address these fundamental limitations, researchers have increasingly focused on developing advanced Non-Fourier heat conduction models. These models incorporate phase lag terms explicitly designed to account for the delayed response between applied thermal gradients, resulting heat flux, and the associated temperature gradients, thus providing a more physically realistic and accurate representation of heat transfer processes in biological tissues [2, 12].

One notable Non-Fourier model extensively employed in bioheat transfer research is the Dual-Phase Lag (DPL) model, which incorporates two distinct phase lags: the heat flux phase lag ($\tau_q$) and the temperature gradient phase lag ($\tau_\theta$) [3, 9, 11, 13]. The heat flux phase lag, $\tau_q$, represents the delayed response of heat flux in reaction to an applied temperature gradient. Physically, this delay arises due to inertia-related phenomena within biological materials, reflecting the finite response time required by heat carriers (such as phonons and electrons) to adjust to rapid changes in the thermal state of the tissue [12, 14]. This inertia causes a measurable delay between the application of a thermal gradient and the resultant initiation of heat flow. On the other hand, the temperature gradient phase lag, $\tau_\theta$, captures the finite time delay in the formation of temperature gradients in response to a changing heat flux. This delay is attributed to complex microstructural interactions occurring at the molecular and microscopic scales, including phonon scattering events, molecular-scale energy exchanges, and thermal relaxation processes inherent to biological tissues [15–17]. Such microscopic couplings contribute significantly to non-instantaneous heat conduction and temperature propagation within tissues. The introduction of these phase lags in the DPL model markedly improves the accuracy and physical representativeness of bioheat transfer simulations. This enhanced accuracy is especially critical in thermal therapy applications such as laser ablation, hyperthermia treatments, and cryosurgery where precise temperature control is necessary to maximize therapeutic effectiveness while ensuring patient safety [8, 13, 17]. The flexibility inherent to the DPL framework further allows researchers to capture diverse thermal behaviors by adjusting the relationship between $\tau_q$ and $\tau_\theta$. Specifically, the model can represent gradient-precedence behavior ($\tau_q > \tau_\theta$), indicating that heat flux lags behind temperature gradients, or flux-precedence behavior ($\tau_q < \tau_\theta$), where temperature gradients lag behind heat flux, thus accommodating a wide range of realistic biological and therapeutic scenarios [3, 16].

Further refinement in modeling bioheat transfer phenomena is accomplished through the Three-Phase Lag (TPL) model, which introduces an additional, distinct phase lag, $\tau_v$, associated explicitly with the thermal displacement gradient [11, 18, 19]. In this advanced modeling approach, thermal displacement ($v$) is explicitly defined so that its time derivative equals the lo-cal temperature field ($\partial v/\partial t = \theta$), effectively linking temperature changes directly to a quantifiable thermal displacement [20]. The additional phase lag, $\tau_v$, specifically captures the finite delay time required for the gradient of this thermal displacement to form in response to a changing heat flux. This concept acknowledges that changes in heat flux do not instantaneously produce corresponding changes in thermal displacement gradients within biological tissues. The significance of including the thermal displacement gradient lag, $\tau_v$, is particularly notable in capturing and accurately representing wave-like thermal conduction phenomena at microscale levels, which are common in biological tissues and other microstructural materials [21, 22]. Such wave-like thermal conduction effects are analogous to mechanical vibrations and manifest clearly during rapid or instantaneous thermal events, for instance, during pulsed laser heating or similarly rapid thermal exposures. These wave-like effects influence the temporal and spatial temperature distributions significantly, causing temperature oscillations and variations that classical models would overlook [22]. By explicitly incorporating $\tau_v$ into the modeling framework, the TPL model not only achieves more physically realistic predictions but also provides smoother and more accurate representations of temperature variations throughout the tissue. This refinement is critically valuable for optimizing therapeutic treatment parameters and enhancing the safety and efficacy of thermal therapies. Additionally, the TPL model provides deeper insights into the microscale mechanisms of heat transport, enabling better understanding and control of bioheat processes in both clinical and experimental contexts [22, 23].

Overall, the adoption of Non-Fourier modeling, through phase lag incorporation, has significantly advanced our understanding of bioheat transfer, material science, and thermal management. By explicitly representing delayed and wave-like thermal responses, these models provide more realistic and precise simulations of medical procedures, including laser irradiation therapies, focused ultrasound treatments, and cryosurgery [7, 24–26]. These enhanced simulations lead directly to improved treatment accuracy, reduced collateral tissue damage, and elevate patient safety standards, making them invaluable tools in clinical practice. As research continues to evolve, ongoing refinements in tissue-specific modeling and the further optimization of Non-Fourier approaches are anticipated. These developments will likely enhance the precision and clinical effectiveness of bioheat modeling, further advancing medical applications and thermal management technologies. Such advancements promise substantial improvements in both therapeutic outcomes and our fundamental understanding of thermal interactions in complex biological systems [8, 13, 22].

The finite difference method is evaluated for modeling TPL heat transfer considering the phase lag terms and for the determination of phase lag terms in bovine skin tissue. The central difference scheme is utilized for spatial discretization and the forward difference scheme is used for temporal discretization to construct the model. The phase lag terms responsible for determining the heat transfer behavior are analyzed using both numerical simulation and analytical mathematics. As in the physical problem, which is three-dimensional, a cylindrical two-dimensional axisymmetric representation is used for a well-founded approximation of the full three-dimensional thermal behavior. By assuming that the reduced model is a great representative for the original system, we enable the reduced model to reflect the majority of characteristics of the original system yet be computationally cheap. To validate the model, 450 nm laser is irradiated to the bovine skin tissue, and the thermal response was collected and recorded by three sensors at different distances from the irradiation site. This data is essential for comparison with the numerical predictions and evaluation of the ability of the analytical approach to characterize phase lag effects in biological tissues. Fig. 1 shows the schematic representation of the modeling setup, with sensor placement and laser irradiation region.

## 2. Mathematical Modeling

According to Fig. 1, the modeling approach implemented in this study is two-dimensional, delineated in the axial ($z$) and also radial ($r$) directions. The origin of the coordinate system is conveniently taken as the center of irradiation since the laser irradiation on the skin tissue is circular shaped and does not deform in the radial direction. By leveraging the inherent symmetry of the three-dimensional physical geometry of the system, due to the symmetry along the third direction, a two-dimensional modeling scenario is adequate for precise simulation [27–29].

$$-\left(\frac{q_r}{r} + \frac{\partial q_r}{\partial r} + \frac{\partial q_z}{\partial z}\right) + Q_L(r, z, t) = \rho c \frac{\partial \theta}{\partial t}. \tag{1}$$

Here, $q_r$ and $q_z$ denote the radial and axial components of the heat flux, respectively. $Q_L$ is the energy flux for the laser irradiation on the tissue.

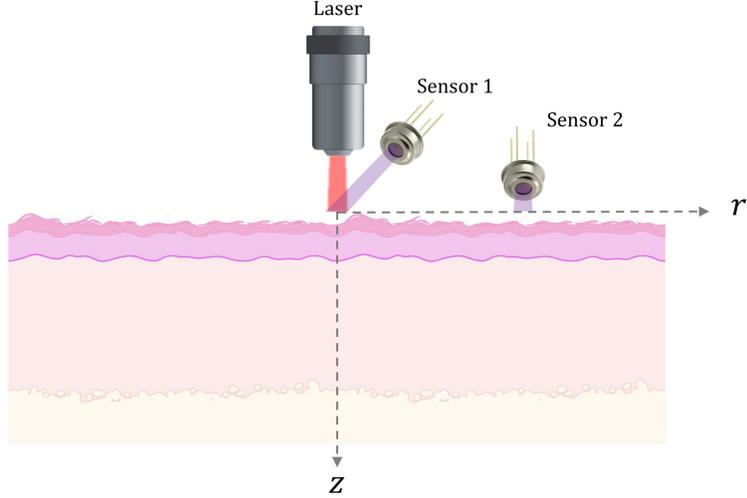

Figure 1: Two-dimensional axisymmetric schematic representation of the skin tissue model subjected to laser irradiation, illustrating the sensor placements and the laser irradiation region. Figure created with BioRender.com.

In addition to this, $\rho$ and $c$ represent density and specific heat capacity, respectively. Additionally, the temperature difference parameter $\theta$ in the governing equations is defined as the change in the local tissue temperature relative to its initial baseline level, written out mathematically as

$$\theta = T_{tissue} - T_0. \tag{2}$$

There are a few investigations that show the models based on DPL and TPL potentially provide better modeling of heat transfer on biological tissues compared to the classic heat conduction models. Although many academic studies have shown that these enhanced models work, there are no definitive conclusions proving one better than others and the phase lag parameters associated with them have not been defined. The governing constitutive equation for the TPL model, as presented in Eq. (3) [30] is used as the last operation. Because of the existing uncertainties on the numerical values of the phase lag terms in simulation and their corresponding interpretation, those parameters have been considered negligible (i.e., zero) during all the modeling.

$$\left(1+\tau_q\frac{\partial}{\partial t}\right)q_i = -k\left(1+\tau_\theta\frac{\partial}{\partial t}\right)\theta_{,i} - k^*\left(1+\tau_v\frac{\partial}{\partial t}\right)v_{,i} \quad ; \quad \dot{v}=\theta. \quad (3)$$

In the Eq. (3) model, four key parameters govern the non-Fourier heat conduction behavior in biological tissues. These include: the heat flux phase lag $\tau_q$, which characterizes the delay in heat flux response to temperature gradients; the temperature gradient phase lag $\tau_\theta$, representing the delay in temperature gradient formation relative to heat flux; the thermal displacement coefficient $k^*$, associated with the wave-like nature of heat conduction; and the thermal displacement phase lag $\tau_v$, which accounts for the delay in the development of thermal displacement gradients. These parameters collectively enable a more accurate depiction of thermal wave propagation and non-equilibrium energy transfer in skin tissue.

The thermal phase lag in this heat transfer progression is the delay from when the heat is provided and when the biological tissue responds to the delivered thermal energy. Skin consists of multiple different types of tissues, some with diverging thermal and micro-structural properties, which causes the delay. These relaxation times or phase delays look very much like inertial phenomenon: they modulate how coils of tissue respond compared to the externally applied heat. Consequently, micro-models representing biological tissues at a cellular level works well for advanced heat conduction theory, whether Fourier or non-Fourier, that incorporates this finite propagation speed and micro-structural effects. As such, an accurate quantification and interpretation of the phase lag responses is necessary for the successful clinical realization of laser-assisted biological procedures, facilitating targeted thermal ablation of pathological structures and minimizing undesired thermal damage upon adjacent healthy tissues [31, 32].

The last term that appears in the constitutive heat conduction equation is due to the Green-Naghdi Type III theory, that takes into account the influence of thermal displacement. This novel theory completes the standard heat conduction theories from classical to dispersion, and more accurately describes wave propagation of heat transfer at finite speed. Specifically, the formulation employed in this work extends the original formulation for the TPL model, which was discussed by Tzou [33], including thermal displacement and a related phase lag term according to Podio-Guidugli et al. [34]. As a result, this more general model allows for better characterization of non-Fourier thermal wave behavior in biological tissues.

By introducing the constitutive equation to the general energy balance equation, the complete governing equation expressing the heat transfer process in the targeted skin tissue system is derived as in eq. $\mathscr{R}_1$:

$$\mathscr{R}_1\left(\theta(r,z,t), Q_L(r,z,t)\right) =$$
$$k\left[\left(\frac{1}{r}\frac{\partial\theta}{\partial r} + \frac{\partial^2\theta}{\partial r^2} + \frac{\partial^2\theta}{\partial z^2}\right) + \tau_\theta\frac{\partial}{\partial t}\left(\frac{1}{r}\frac{\partial\theta}{\partial r} + \frac{\partial^2\theta}{\partial r^2} + \frac{\partial^2\theta}{\partial z^2}\right)\right] +$$
$$k^*\left[\left(\frac{1}{r}\frac{\partial v}{\partial r} + \frac{\partial^2 v}{\partial r^2} + \frac{\partial^2 v}{\partial z^2}\right) + \tau_v\frac{\partial}{\partial t}\left(\frac{1}{r}\frac{\partial v}{\partial r} + \frac{\partial^2 v}{\partial r^2} + \frac{\partial^2 v}{\partial z^2}\right)\right] + \quad (4)$$
$$\left[Q_L(r,z,t) - \rho c\frac{\partial\theta}{\partial t}\right] + \tau_q\left[\frac{\partial Q_L(r,z,t)}{\partial t} - \rho c\frac{\partial^2\theta}{\partial t^2}\right] = 0 \,;$$
$$v = \int_0^t \theta(r,z,\tau)d\tau$$

The Beer-Lambert law describes the basic interactions occurring when radiation is absorbed in a medium and thus is important for modeling the radiation propagation in biological tissues like skin. Such a notion quantitatively governs the exponential loss of the radiation intensity while traversing through a radiation-absorbing media, providing a sound basis for measuring the energy deposition profiles in laser-exposed tissue. This law provides a universal approach for quantifying radiation absorption processes across multiple fields of science and is therefore frequently utilized. Therefore, such law is a fundamental characteristic aspect describing optical traits and thermal behaviors across biological tissues and, as such contribute to the establishment and optimization of laser-based therapeutic strategies [30, 35–38]. Considering the Gaussian spatial distribution, the total laser power is computed by integrating the laser intensity over the radial axis. This integration creates a mathematical link between the total laser power and the peak intensity, $I_0^{(P)}$. The developed relationship is expressed as follows:

$$\int_0^\infty 2\pi r I_0^{(P)} e^{\left(\frac{-r^2}{2\sigma^2}\right)} dr = P; \, I_0^{(P)} = \frac{P}{2\pi\sigma^2}. \quad (5)$$

Equation (5) shows the exponential attenuation of laser beam intensity as it travels through the medium, a process influenced by absorption and scattering events. The optical characteristics of the material especially the absorption coefficient and the scattering coefficient determine the attenuation, which together govern the rate of laser energy loss. The thermal power density absorbed by the composite directly relates to the variation in laser intensity at two separate depths into the medium. Facilitating thermal phenomena including heat conduction and localized heating inside the material requires the absorbed energy. Mathematically, the absorbed thermal power density in the composite is the spatial rate of change of laser intensity with respect to the depth coordinate. As shown in Equation (6), the relationship is represented by the derivative of laser intensity with respect to the depth axis. This approach defines a distinct link between the thermal reaction of the material and the dispersion of laser energy.

$$Q_L = \frac{P}{2\pi\sigma^2} e^{\left(\frac{-r^2}{2\sigma^2}\right)} \mu_a e^{-\mu_a z}. \tag{6}$$

The first condition prescribes no heat flows across the boundary at $r = 0$, which is a particle-symmetry in the rest frame of the particle. This condition is mathematically expressed by Eq. (7)

$$q_r\big|_{r=0} = 0 \quad ; \quad \frac{\partial \theta(r,z,t)}{\partial r}\bigg|_{r=0} = 0. \tag{7}$$

Moreover, the thermal convection from the skin's surface at $z = 0$ must equal the thermal conduction across that surface, which can be expressed as:

$$-k\frac{\partial \theta(r,z,t)}{\partial z}\bigg|_{z=0} - h\Theta(r,0,t) = 0. \tag{8}$$

The remaining two boundary conditions, which are straightforward, arise as $r \to \infty$ and $z \to \infty$. At these two boundaries, the temperature rise approaches zero. As $z$ approaches infinity, laser power penetration is nonexistent, resulting in an insignificant temperature increase. Likewise, as $r \to \infty$, radial heat transfer cannot induce a temperature elevation, resulting in negligible temperature variation.

$$\lim_{z\to\infty} \theta(r,z,t) = 0 \quad ; \quad \lim_{r\to\infty} \theta(r,z,t) = 0 \tag{9}$$

But one proceeds as if the system obeys constitutive Eq. (3), the new equation for defining boundary condition represented by $\mathscr{R}_2$, can be obtained. This new equation is called Eq. (10)

$$\mathscr{R}_2\left(\theta(r,0,t)\right) = -k\left(1+\tau_\theta\frac{\partial}{\partial t}\right)\frac{\partial\theta}{\partial z}\bigg|_{z=0} \\ -k^*\left(1+\tau_v\frac{\partial}{\partial t}\right)\frac{\partial v}{\partial z}\bigg|_{z=0} = h\left(1+\tau_q\frac{\partial}{\partial t}\right)\left[\theta(0,t)-\theta_{amb}\right], \quad (10)$$

where $h$ is the convection coefficient between the skin tissue and air, and $\theta_{amb}$ is the ambient temperature with which the tissue has thermal convection.

## 3. Experimental Setup and Simulation Parameters

The realization of this experiment for bovine skin was directed towards measurement of the phase lag values of the bovine skin tissue of the same and results interpreted to measure the tissue phase properties under controlled laser heat source. The configuration consisted of a high-intensity laser source and non-contact infrared temperature sensors, as well as shielding components to facilitate accurate thermal measurements. For this experiment, bovine skin tissue was selected as it most closely mimicked both structural and thermal properties of human skin. A heat source of 450 nm and 4000 mW laser source was used. The wavelength range is in the visible spectrum, therefore penetrating to a low depth in biological tissue and where most of the energy is absorbed at the skin surface. As the energy is absorbed, localized heating occurs, which allows the surface temperature changes to be precisely analyzed for a non-Fourier heat transfer investigation. Three MLX90614 non-contact infrared temperature sensors have been used to register the thermal response of the tissue. These sensors are used because of their accuracy and quick response times. The MLX90614 sensors have resolution of 0.05 K, accuracy of less than 0.5 K, and sampling interval of 0.1 seconds. Because these sensors take readings from an area, instead of a single point, they will output a temperature value weighted based on their Field Of View (FOV) function.

The temperature sensors were placed in a radial manner to capture the temperature distribution. Sensors were placed in the following configurations as shown in Figure 1 where Sensor 1 was situated along the same axis as the laser irradiation point to capture the maximum temperature response. Sensor 2 was placed at equal distance radially away from the laser irradiation site to monitor heat propagation through the tissue. To avoid direct laser interference with the sensors, metal shielding was placed between the laser beam and the sensors. This shields suitably scattered the laser light without obstructing the sensors recording the tissue temperature and there were no optical artifacts. Here the fresh bovine skin sample was placed on the surface thermally insulated to avoid losses of heat from surroundings. The 450 nm laser was turned on and pointed to a predetermined position on the tissue surface, during which time the three MLX90614 sensors constantly recorded temperature changes at their positions every 0.1 second. The measured temperature profiles were analyzed to assess the phase lag behavior and verify the non-Fourier heat conduction model.

The experiment involved some assumptions and considerations. Assuming thermal homogeneity of the tissue sample in the measurement area, thermal convection and evaporation heat loss effects were neglected during the current laser exposure duration. The metal cover did an effective job of stopping direct laser irradiation from affecting the sensor measurements. The present configuration was such that accurate data collection could take place to specifically validate TPL heat conduction model in bio tissue. The obtained temperature data were then utilized to obtain phase lag values and validate numerical model predictions.

The selected biological tissue for this work is bovine skin samples obtained circularly from the lower leg area of the animal, as illustrated in Fig. 2. Generated for experimental homogeneity were several samples. To avoid dehydration and maintain the tissue's natural moisture content, all of the samples were kept in deionized water all through the testing process. Extra water was painstakingly eliminated to ensure a dry surface before laser irradiation since the immersion in deionized water was supposed to just prevent moisture loss rather than alter the thermal properties of the tissue.

As already seen in Fig. 1, three infrared temperature sensors were positioned purposefully to capture the temperature dispersion in the tissue samples. The first sensor was positioned exactly at the point of laser irradiation to capture the maximum temperature response. The second and third sensors were positioned respectively at radial distances of 1 cm and 2 cm from the center of the irradiation site in order to monitor the spatial propagation of heat across the tissue. The 3 mm radius laser beam assured a precisely defined irradiation region. Dermal and subcutaneous fat layers abound in every 1 cm thick, 6 cm diameter excised cow skin sample. The tissue specimens were precisely cut in circular forms to maintain geometric symmetry. This symmetry assured that the laser beam always incident at the center of the sample, so preserving coordinate system consistency. This approach allowed the implementation of a two-dimensional axisymmetric heat transport model, therefore providing a computationally efficient approximation of the thermal response in the tissue.

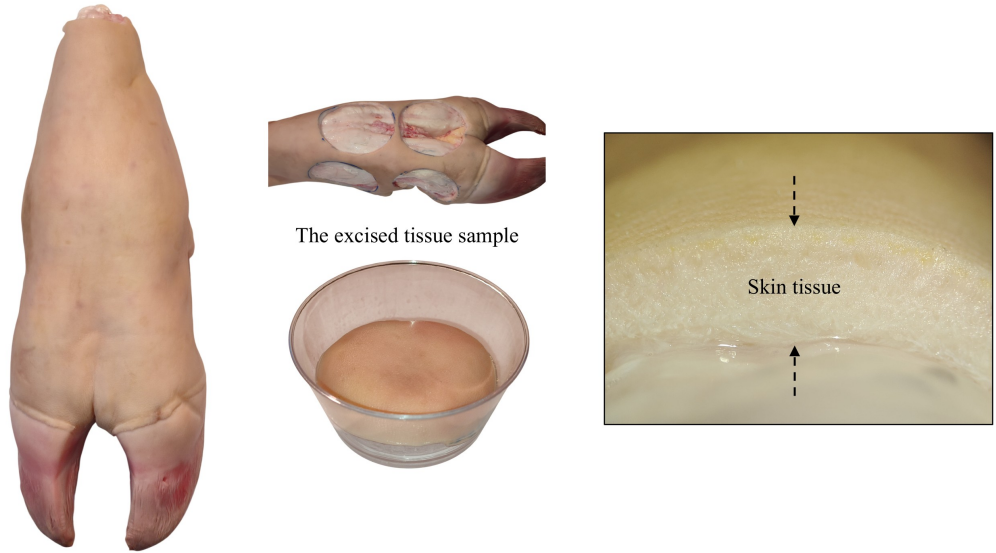

Figure 2: The sample of bull skin tissue was used, and the cut section of it was prepared for laser irradiation tests.

Moreover, according to the analysis of the body of current research by the writers, no thorough experimental studies have been carried out to exactly ascertain the thermal characteristics of bovine skin. Hence, no particular reference exists that offers precise numerical values for these characteristics. Nevertheless, an estimate of the thermal characteristics needed for the simulation has been found by aggregating data from several sources. Table 1 summarizes the particular thermal values used in the modeling method.

Simulations were also carried out utilizing a range of various thermal conductivity values to guarantee that uncertainty in the thermal property values does not compromise the accuracy of the derived phase lag estimates. Discussed in next parts, this sensitivity study enables the evaluation of possible phase lag value deviations resulting from changes in the anticipated thermal

Table 1: Thermal properties of bovine skin used in the simulation

| Property | Symbol | Value |
|---|---|---|
| Thermal Conductivity (W/m · K) | $k$ | 0.45 [39, 40] |
| Specific Heat Capacity (J/kg · K) | $c$ | 3500 [41] |
| Density (kg/m³) | $\rho$ | 1000 [42] |
| Convection Coefficient (W/m² · K) | $h$ | 15 [43] |

parameters.

*3.1. Cow Skin Tissue: A Substitute for Human Skin in Experiments*

Human skin and cow skin possess a similar full-thickness structure and composition, which leads to comparable thermal behaviors. Both tissues have an outer epidermis and a thicker, collagen rich dermis, with high water content and extensive collagen fiber networks [44]. These common material characteristics play a crucial role in determining their heat transfer and energy storage capabilities. In the following, key thermal properties namely, thermal conductivity, specific heat capacity, and thermal diffusivity are compared for human skin and cow skin based on experimental data and engineering material values.

Table 2: Comparison of Full Thickness Human Skin Thermal Properties

| Thermal Property | Human Skin |
|---|---|
| Thermal Conductivity (W/m·K) | 0.32 to 0.50 [30, 45, 46] |
| Specific Heat Capacity (J/kg·K) | ≈ 3400 [27, 28, 47] |

In summary, the thermal profiles of human and cow skin are similar. Both exhibit moderate thermal conductivity values (typically in the range of 0.32 to 0.50 W/m·K for human skin and approximately 0.30 W/m·K for hydrated cow skin) and high specific heat capacities (around 3.5 kJ/kg·K), largely due to their substantial water and collagen content. Their thermal diffusivities, approximately 0.10 mm$^2$/s, indicate that both tissues respond similarly to temperature changes. Thus, cow skin serves as an effective analog for human skin in thermal experiments, reliably simulating the heat transfer characteristics observed in human tissue for engineering and biomedical research.

## 4. Numerical Method

Implementing the implicit finite difference method (central difference method for spatial discretization and forward difference method for temporal discretization) has resulted in the numerical solution of the governing equations. This numerical approach is well known from its superior accuracy and numerical stability for simulation of complex thermal interactions when biological tissues become exposed to laser irradiation [48–52]. The discretized governing heat transfer equation employed in this study is expressed by Eq. (11).

$$\frac{\partial \theta}{\partial r} \approx \frac{\theta^{l}_{(m+1,n)} - \theta^{l}_{(m-1,n)}}{2\Delta r}, \qquad \frac{\partial^{2}\theta}{\partial r^{2}} \approx \frac{\theta^{l}_{(m+1,n)} - 2\theta^{l}_{(m,n)} + \theta^{l}_{(m-1,n)}}{\Delta r^{2}},$$

$$\frac{\partial^{2}\theta}{\partial z^{2}} \approx \frac{\theta^{l}_{(m,n+1)} - 2\theta^{l}_{(m,n)} + \theta^{l}_{(m,n-1)}}{\Delta z^{2}},$$

$$\frac{\partial}{\partial t}\left(\frac{\partial \theta}{\partial r}\right) \approx \frac{\theta^{f+1}_{(m+1,n)} - \theta^{f+1}_{(m-1,n)} - \theta^{l}_{(m+1,n)} + \theta^{l}_{(m-1,n)}}{2\Delta r \Delta t},$$

$$\frac{\partial}{\partial t}\left(\frac{\partial^{2}\theta}{\partial r^{2}}\right) \approx \frac{\theta^{f+1}_{(m+1,n)} - 2\theta^{f+1}_{(m,n)} + \theta^{f+1}_{(m-1,n)} - \theta^{l}_{(m+1,n)} + 2\theta^{l}_{(m,n)} - \theta^{l}_{(m-1,n)}}{\Delta r^{2} \Delta t},$$

$$\frac{\partial}{\partial t}\left(\frac{\partial^{2}\theta}{\partial z^{2}}\right) \approx \frac{\theta^{f+1}_{(m,n+1)} - 2\theta^{f+1}_{(m,n)} + \theta^{f+1}_{(m,n-1)} - \theta^{l}_{(m,n+1)} + 2\theta^{l}_{(m,n)} - \theta^{l}_{(m,n-1)}}{\Delta z^{2} \Delta t},$$

$$\frac{\partial v}{\partial r} = \int_{0}^{t} \frac{\partial \theta}{\partial r} \approx \sum_{l=1}^{\lfloor \frac{t}{\Delta t}\rfloor} \frac{\theta^{l}_{(m+1,n)} - \theta^{l}_{(m-1,n)}}{2\,r}\Delta t,$$

$$\frac{\partial^{2} v}{\partial r^{2}} = \int_{0}^{t} \frac{\partial^{2}\theta}{\partial r^{2}} \approx \sum_{l=1}^{\lfloor \frac{t}{\Delta t}\rfloor} \frac{\theta^{l}_{(m+1,n)} - 2\theta^{l}_{(m,n)} + \theta^{l}_{(m-1,n)}}{\Delta r^{2}}\Delta t,$$

$$\frac{\partial^{2} v}{\partial z^{2}} = \int_{0}^{t} \frac{\partial^{2}\theta}{\partial z^{2}} \approx \sum_{l=1}^{\lfloor \frac{t}{\Delta t}\rfloor} \frac{\theta^{l}_{(m,n+1)} - 2\theta^{l}_{(m,n)} + \theta^{l}_{(m,n-1)}}{\Delta z^{2}}\Delta t, \qquad \frac{\partial Q_{L}(r,z,t)}{\partial t} \approx \frac{Q_{L(m,n)}^{f+1} - Q_{L(m,n)}^{l}}{\Delta t},$$

$$\frac{\partial \theta}{\partial t} \approx \frac{\theta^{f+1}_{(m,n)} - \theta^{l}_{(m,n)}}{\Delta t}, \qquad \frac{\partial^{2}\theta}{\partial t^{2}} \approx \frac{\theta^{l+2}_{(m,n)} - 2\theta^{f+1}_{(m,n)} + \theta^{l}_{m,n}}{\Delta t^{2}},$$

$$\frac{\partial}{\partial t}\left(\frac{\partial v}{\partial r}\right) = \frac{\partial \theta}{\partial r} \approx \frac{\theta^{l}_{(m+1,n)} - \theta^{l}_{(m-1,n)}}{2\Delta r}, \qquad \frac{\partial}{\partial t}\left(\frac{\partial^{2} v}{\partial r^{2}}\right) = \frac{\partial^{2}\theta}{\partial r^{2}} \approx \frac{\theta^{l}_{(m+1,n)} - 2\theta^{l}_{(m,n)} + \theta^{l}_{(m-1,n)}}{\Delta r^{2}},$$

$$\frac{\partial}{\partial t}\left(\frac{\partial^{2} v}{\partial z^{2}}\right) = \frac{\partial^{2}\theta}{\partial z^{2}} \approx \frac{\theta^{l}_{(m,n+1)} - 2\theta^{l}_{(m,n)} + \theta^{l}_{(m,n-1)}}{\Delta z^{2}}.$$

(11)

By substituting the indicated terms in Eq. (11) into $\mathscr{R}_1$ (Eq. (4)) and rewriting the equation as shown below, the governing equation for the sys-tem can be solved using the implicit central method. Rearranging this equation with respect to the term $\tau_q \rho c \frac{\theta^{l+2}_{(m,n)}}{\Delta t^2}$ yields a structured form suitable for numerical solution using the finite difference method.

$$k \left[ \left( \frac{1}{m\Delta r} \frac{\theta^l_{(m+1,n)} - \theta^l_{(m-1,n)}}{2\Delta r} + \frac{\theta^l_{(m+1,n)} - 2\theta^l_{(m,n)} + \theta^l_{(m-1,n)}}{\Delta r^2} + \frac{\theta^l_{(m,n+1)} - 2\theta^l_{(m,n)} + \theta^l_{(m,n-1)}}{\Delta z^2} \right) \right.$$

$$+ \tau_\theta \left( \frac{1}{m\Delta r} \frac{\theta^{l+1}_{(m+1,n)} - \theta^{l+1}_{(m-1,n)} - \theta^l_{(m+1,n)} + \theta^l_{(m-1,n)}}{2\Delta r \Delta t} \right.$$

$$+ \frac{\theta^{l+1}_{(m+1,n)} - 2\theta^{l+1}_{(m,n)} + \theta^{l+1}_{(m-1,n)} - \theta^l_{(m+1,n)} + 2\theta^l_{(m,n)} - \theta^l_{(m-1,n)}}{\Delta r^2 \Delta t}$$

$$\left. \left. + \frac{\theta^{l+1}_{(m,n+1)} - 2\theta^{l+1}_{(m,n)} + \theta^{l+1}_{(m,n-1)} - \theta^l_{(m,n+1)} + 2\theta^l_{(m,n)} - \theta^l_{(m,n-1)}}{\Delta z^2 \Delta t} \right) \right]$$

$$+ k^* \left[ \left( \frac{1}{m\Delta r} \sum_{l=1}^{\lfloor \frac{t}{\Delta t} \rfloor} \frac{\theta^l_{(m+1,n)} - \theta^l_{(m-1,n)}}{2\Delta r} \Delta t + \sum_{l=1}^{\lfloor \frac{t}{\Delta t} \rfloor} \frac{\theta^l_{(m+1,n)} - 2\theta^l_{(m,n)} + \theta^l_{(m-1,n)}}{\Delta r^2} \Delta t \right. \right.$$

$$\left. + \sum_{l=1}^{\lfloor \frac{t}{\Delta t} \rfloor} \frac{\theta^l_{(m,n+1)} - 2\theta^l_{(m,n)} + \theta^l_{(m,n-1)}}{\Delta z^2} \Delta t \right)$$

$$\left. + \tau_v \left( \frac{1}{m\Delta r} \frac{\theta^l_{(m+1,n)} - \theta^l_{(m-1,n)}}{2\Delta r} + \frac{\theta^l_{(m+1,n)} - 2\theta^l_{(m,n)} + \theta^l_{(m-1,n)}}{\Delta r^2} + \frac{\theta^l_{(m,n+1)} - 2\theta^l_{(m,n)} + \theta^l_{(m,n-1)}}{\Delta z^2} \right) \right]$$

$$+ \left[ Q^l_{L(m,n)} - \rho c \frac{\theta^{l+1}_{(m,n)} - \theta^l_{(m,n)}}{\Delta t} \right] + \tau_q \left[ \frac{Q^{l+1}_{L(m,n)} - Q^l_{L(m,n)}}{\Delta t} - \rho c \frac{\theta^{l+2}_{(m,n)} - 2\theta^{l+1}_{(m,n)} + \theta^l_{(m,n)}}{\Delta t^2} \right] = 0 \quad (12)$$

## 5. Numerical results

Figure 3 illustrates the baseline simulation setup using the Fourier heat conduction model, with all phase-lag coefficients set to zero. In this figure, the temperature distribution is plotted in three dimensions over the *r* and *z*

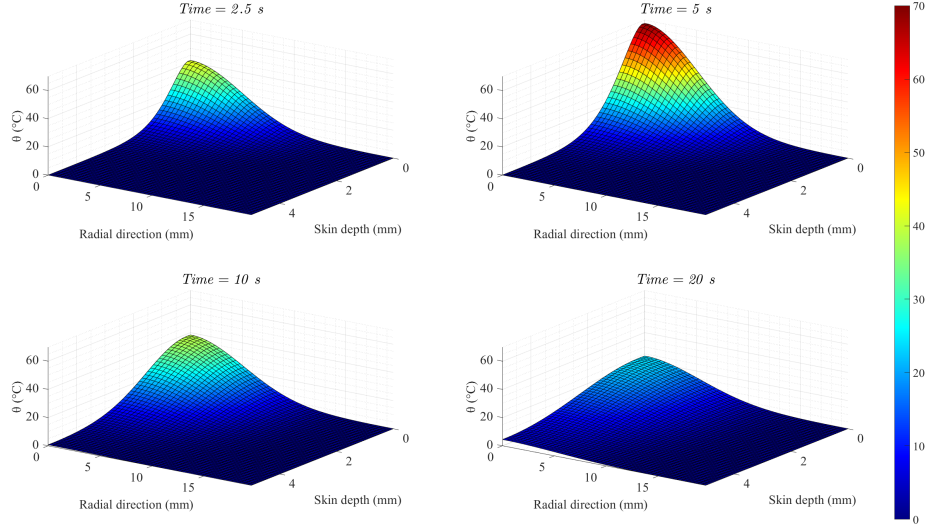

Figure 3: Two-dimensional simulation results showing temperature distribution profiles across skin depth and radial directions at selected time intervals. ($\tau_q = 0.01$, $\tau_\theta = 0$, $\tau_v = 0$, $k^* = 0$)

directions at four time instants, namely 2.5 s, 5 s, 10 s, and 20 s. The laser irradiation is active from the beginning of the simulation until $t = 5$ s.

Although these 3D visualizations help us grasp the overall spatial and temporal evolution of temperature, it is challenging to conduct precise comparisons or display multiple scenarios side-by-side in this form. Therefore, we will also provide two-dimensional plots at specific *r* or *z* values and time instants in order to perform more detailed analyses and to compare different curves directly. This initial figure serves as a reference case under purely Fourier (classical) assumptions, setting the stage for later discussions where non-Fourier models and the associated phase-lag parameters are introduced and investigated.

## 5.1. Analysis of $\tau_q$

Figure 4 illustrates the time–temperature profiles at four distinct locations within the skin tissue, covering both the heating period (from $t = 0s$ to $t = 5s$ during laser irradiation) and the subsequent cooling period (from $t = 5s$ to $t = 20s$ after the laser is turned off). In this simulation, all other phase lag parameters are set to zero, and only $\tau_q$ varies; therefore, this model corresponds to a SPL formulation.

The results clearly show that increasing $\tau_q$ intensifies both the heating and cooling slopes. In other words, a larger $\tau_q$ enhances the heating rate during irradiation and accelerates the cooling rate once the laser is off. This behavior is visible in all four subplots.

In cases where $\tau_q$ is large, the sharp thermal transitions lead to instability at some locations. Specifically, for certain spatial points, heat is rapidly introduced from one side and simultaneously exits from the other, which creates sharp overshoots followed by deep undershoots. Such thermal fluctuations are unlikely to be observed in realistic physical systems, suggesting that large $\tau_q$ values may cause non-physical or exaggerated responses in this model.

Moreover, in regions that are not directly exposed to the laser and are instead heated gradually via thermal conduction such as the point at ($r = 15$ mm, $z = 4$ mm) the final temperature rise is smaller when $\tau_q$ is larger. This occurs because, with larger $\tau_q$, the heating process becomes shorter in effect due to faster dissipation once the irradiation ends. As a result, remote regions have less opportunity to accumulate thermal energy before the cooling phase dominates.

In summary, increasing $\tau_q$ steepens both the rise and fall of the temperature response and can cause thermal instability at certain points, especially near and shortly after the heat source. Meanwhile, distant locations that rely solely on conductive heat transfer experience lower peak temperatures as $\tau_q$ increases. These observations underline the transient and sometimes unrealistic effects introduced by large heat flux phase lags in SPL-based thermal models.

Figure 5 presents the same simulation cases previously shown in Figure 4, but instead of showing the temperature over time at fixed spatial points, it shows the temperature distribution along the direction of depth of the skin (from $z = 0$ to $z = 5$ mm) at selected time snapshots. In this view, we can spatially evaluate how the temperature propagates through the depth of the skin at different stages of the heating and cooling cycle.

As also observed in Figure 4, up to $t = 5$ s, the period during which laser irradiation is active cases with higher values of $\tau_q$ reach higher temperatures. This behavior is again evident in the current figure, especially at the surface ($z = 0$), where the thermal energy accumulates more significantly in higher cases $\tau_q$ due to the delayed heat flux response.

Once the laser is turned off at $t = 5$ s, a rapid temperature drop begins,

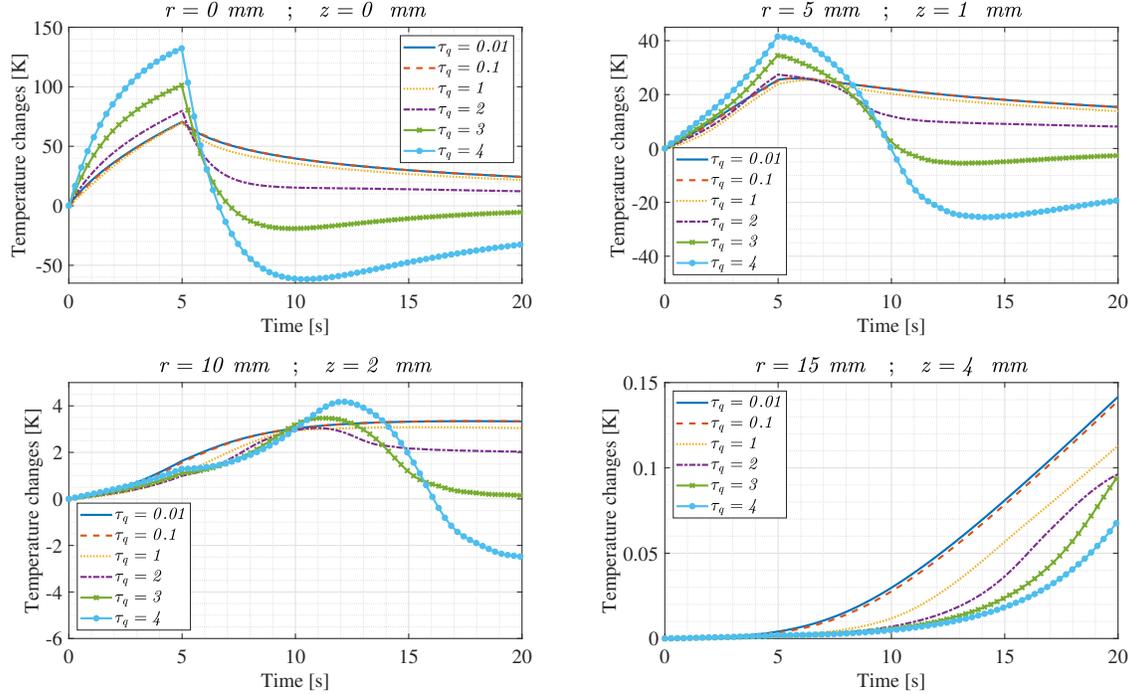

Figure 4: Examining the temperature time history at different locations from the two dimensional modeling platform for different $\tau_q$ values. ($\tau_\theta = 0$, $\tau_v = 0$, $k^* = 0$)

particularly in cases with large $\tau_q$. This drop initiates from the surface and propagates downward. Even if air convection was absent, such a drop would still occur in high $\tau_q$ scenarios due to the sharp reversal of heat flux i.e., heat entering the tissue during irradiation and then leaving it just as rapidly afterward. However, the presence of convective heat loss to the ambient air further amplifies this behavior. It accelerates the cooling process right at the surface, creating a strong negative heat flux that pulls energy out of the tissue. As a result, the steepest temperature gradients are observed near the surface, and a wave-like thermal decay front appears to travel into the tissue.

Notably, after $t = 7$ s, highly non-physical temperature drops are observed in the high $\tau_q$ cases. Temperature reductions of up to 20 to 60 degrees, relative to the initial baseline temperature are observed. Such drastic undershoots are not physically plausible in biological tissue, as it is extremely unlikely that passive cooling mechanisms produce such strong temperature

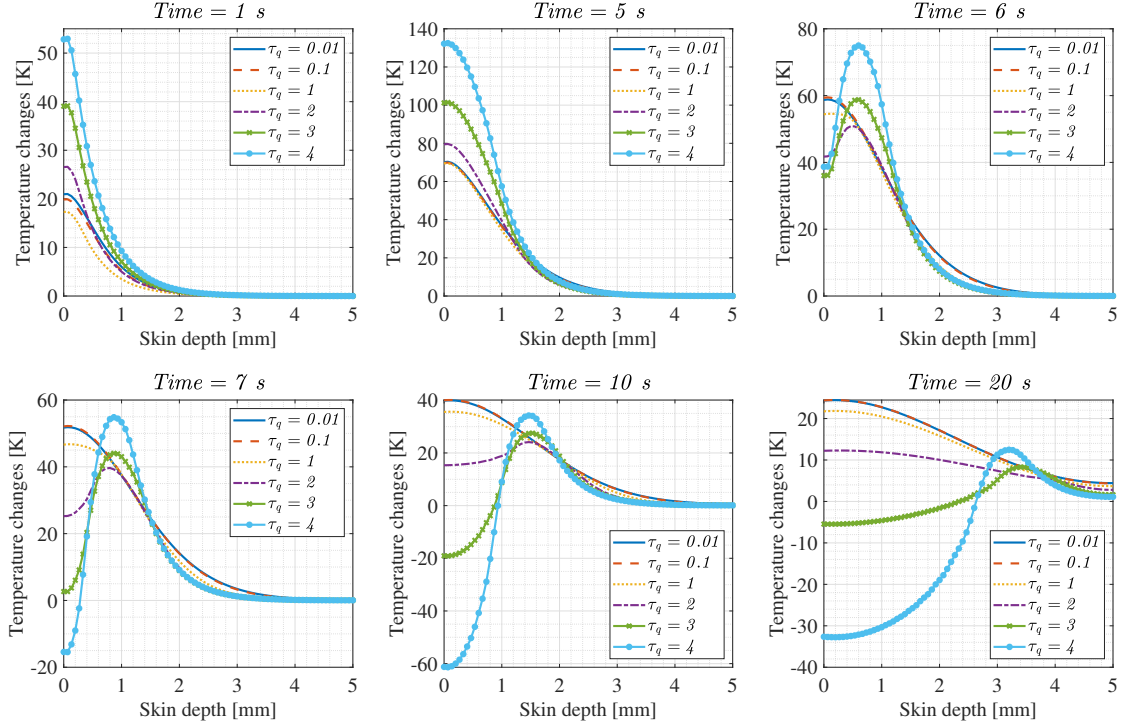

Figure 5: Investigation of temperature profiles in skin tissue depth for different $\tau_q$ values at various time steps. ($r = 0$ mm, $\tau_\theta = 0$, $\tau_v = 0$, $k^* = 0$)

dips within seconds of stopping irradiation.

These unrealistic results reinforce the conclusion from Figure 4 that large $\tau_q$ values can induce thermal instabilities and oscillations that are not representative of real physical behavior. Therefore, the value of $\tau_q$ must be carefully selected within a realistic and experimentally supported range, which will be addressed in later sections.

Figure 6 illustrates the same simulation cases as those shown in Figures 4 and 5, but this time the temperature is plotted along the radial direction ($r = 0$ to $r = 20$ mm) at the skin surface (i.e., at fixed depth $z = 0$). This allows for evaluating the lateral spread of heat across the irradiated surface under different values of the heat flux phase lag $\tau_q$.

Unlike Figure 5, where a wave-like appearance emerged in the temperature-depth profiles due to the rapid cooling effect initiated from the surface (af-fected by convection), here in Figure 6 such wave-shaped profiles are no longer observed. The key reason is that in this case, all plotted points lie on the skin surface, where convective heat loss to the ambient air is applied uniformly. As a result, the entire radial temperature profile evolves more smoothly, and the temperature decay appears to be continuous and monotonic without local dips or crest-like forms. Therefore, the superficial "wave-shape" observed in depthwise views is not present in this radial surface view, despite the same underlying physics.

As shown in earlier figures, during the heating phase (e.g., $t = 1\,\text{s}$ and $t = 5\,\text{s}$), cases with larger $\tau_q$ exhibit significantly higher temperatures near the laser center. This trend continues along the radial profile, with the highest central temperatures corresponding to $\tau_q = 4$, and progressively lower peaks as $\tau_q$ decreases.

Once the laser is turned off at $t = 5\,\text{s}$, a rapid cooling phase begins, especially in cases with large $\tau_q$. From $t = 6\,\text{s}$ onward, the profiles for high values of $\tau_q$ begin to drop sharply. By $t = 7\,\text{s}$, these cases begin to exhibit negative temperature changes (relative to the initial condition). At $t = 10\,\text{s}$, the $\tau_q = 4$ case reaches a minimum of approximately $-60\,\text{K}$, while by $t = 20\,\text{s}$, this same case recovers to around $-30\,\text{K}$. This trend illustrates the exaggerated thermal response associated with large $\tau_q$: an aggressive post-heating drop followed by a gradual rebound as lateral conduction restores the temperature to its baseline.

Overall, Figure 6 confirms that large values of $\tau_q$ in SPL modeling can lead to unrealistic thermal dynamics on the skin surface. Although the surface profiles in this figure do not exhibit the artificial wave-like shapes seen in Figure 5, the excessive cooling depths and strong undershoots up to $60\,\text{K}$ below baseline still reflect the physical limitations of high $\tau_q$ values in biological tissue. These findings further support the need to determine a realistic and physiologically valid range for $\tau_q$, which will be addressed through experimental comparison in later sections.

*5.2. Analysis of $\tau_\theta$*

Figure 7 presents the results of a simulation based on the DPL model, where the heat flux phase lag $\tau_q$ is fixed at 1, and the temperature gradient phase lag $\tau_\theta$ is varied. The time-dependent thermal response is analyzed at four distinct locations in the skin tissue. Laser irradiation is applied until $t = 5\,\text{s}$, after which the system enters a cooling phase up to $t = 20\,\text{s}$.

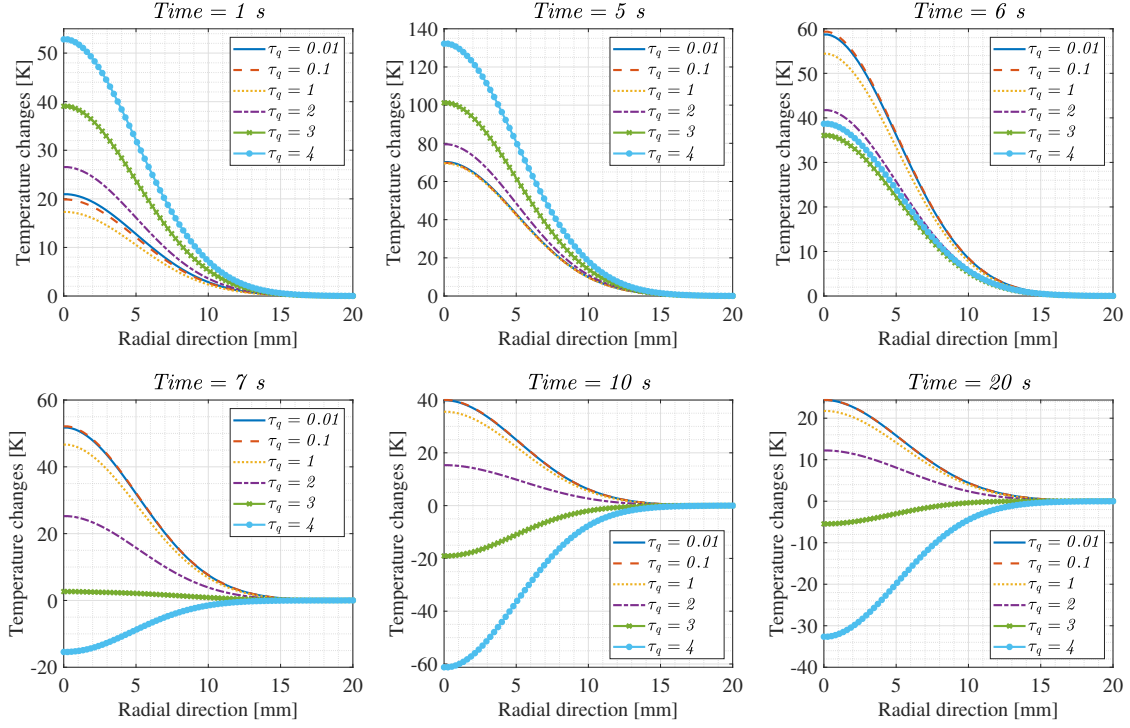

Figure 6: Investigation of temperature profiles in the radial direction of skin tissue for different $\tau_q$ values at various time steps. ($z = 0$ mm, $\tau_\theta = 0$, $\tau_v = 0$, $k^* = 0$)

At the first location ($r = 0$ mm, $z = 0$ mm), which is directly exposed to the laser, increasing $\tau_\theta$ results in lower peak temperatures during irradiation. This occurs because a higher $\tau_\theta$ enhances the effective thermal conductivity in the irradiated zone, allowing the incoming energy to spread outward more quickly, thus limiting the local temperature rise. After the laser is turned off, cases with larger $\tau_\theta$ retain their heat longer and cool down more slowly due to reduced post-irradiation heat conduction. As a result, at $t = 20$ s, the case with $\tau_\theta = 8$ exhibits the highest residual temperature at this location.

At the second location ($r = 5$ mm, $z = 1$ mm), a similar but less pronounced trend is observed. During irradiation, higher $\tau_\theta$ values facilitate faster heat transfer from the source to this point, and after irradiation ends, these cases maintain higher temperatures than those with smaller $\tau_\theta$.

At the third location ($r = 10$ mm, $z = 2$ mm), which lies completely out-side the irradiated area, the temperature is higher for larger $\tau_\theta$ values of up to $t = 5\,\text{s}$. This is because of increased thermal conductivity in the irradiated zone allows heat to reach this point more efficiently. However, after the laser is turned off, larger $\tau_\theta$ leads to decreased conductivity in the irradiated zone, slowing down the delivery of heat to this point. As a result, lower temperatures are observed for high $\tau_\theta$ cases in the cooling phase.

At the fourth location ($r = 15\,\text{mm}$, $z = 4\,\text{mm}$), which is the farthest point from the laser source, the temperature increases continuously in all cases. This region receives heat from the area directly irradiated by the laser. Cases with larger $\tau_\theta$ reach higher temperatures at this location, as thermal energy is transferred more effectively from the center. After the laser is turned off, the temperature continues to rise with nearly the same slope across all cases, but the absolute temperature remains highest for $\tau_\theta = 8$ throughout the remaining simulation period.

Overall, Figure 7 demonstrates that increasing $\tau_\theta$ alters both the timing and magnitude of temperature evolution throughout the tissue, with strong influence on both energy dispersion and retention characteristics.

Figure 8 further supports the findings discussed in Figure 7 and validates the behavior expected from the DPL model. This figure presents the depth-wise temperature distribution from the skin surface down to a depth of 5 mm at multiple time points, under different values of the temperature gradient phase lag $\tau_\theta$.

In the early stages of simulation, such as at 1 s and 5 s, the surface region directly exposed to laser heating shows that higher $\tau_\theta$ values result in lower surface temperatures. This is because a larger $\tau_\theta$ increases the effective thermal conductivity near the laser-irradiated zone, allowing the absorbed energy to quickly spread toward deeper regions. As a result, the surface retains less heat and remains cooler.

However, moving deeper into the tissue at those same early times reveals the opposite trend: higher $\tau_\theta$ cases exhibit higher temperatures in the deeper parts of the skin. This occurs because the increased conductivity in the irradiated zone enables faster heat transfer from the surface to the interior. Thus, regions with greater skin depth absorb more thermal energy and experience elevated temperatures compared to cases with lower $\tau_\theta$.

As time progresses from approximately 6 s to 20 s the differences among the various $\tau_\theta$ cases gradually diminish. This is because the influence of $\tau_\theta$ is most significant when there are steep temperature gradients. In such conditions, the gradient of phase lag strongly affects thermal diffusivity. But over time, as the heat spreads more uniformly, regions with initially higher temperatures lose energy faster, while cooler regions absorb heat more readily. This mutual exchange leads to a relative thermal balance, and the temperature profiles of all cases begin to converge.

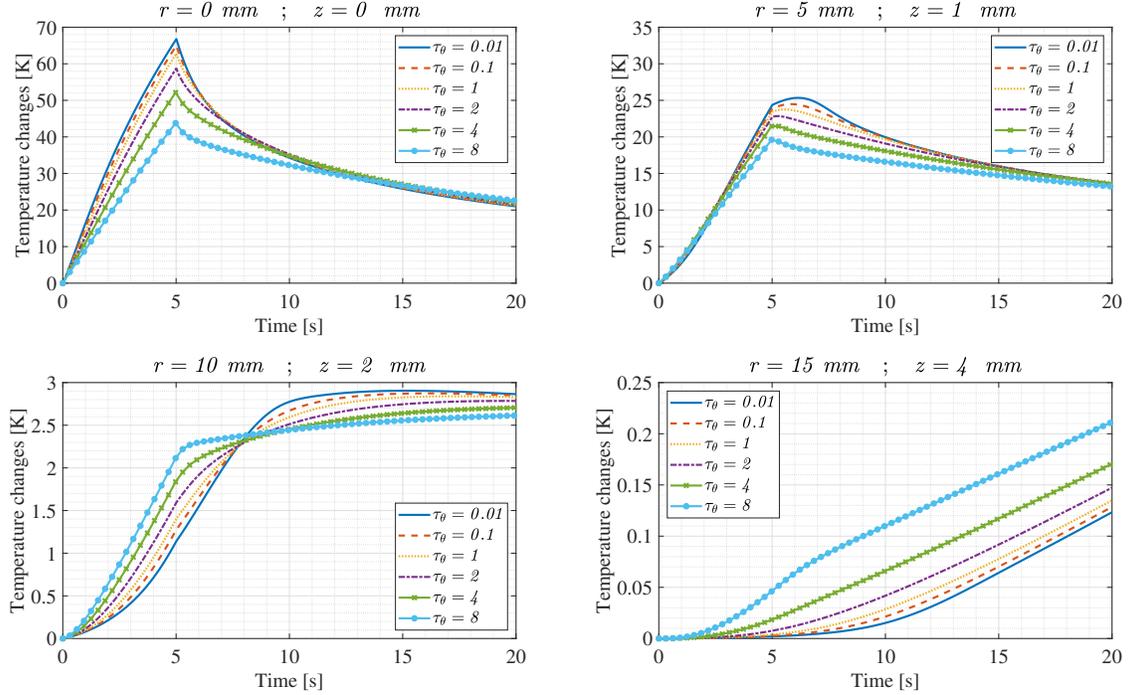

Figure 7: Examining the temperature time history at different locations from the two dimensional modeling platform for different $\tau_\theta$ values. ($\tau_q = 1$, $\tau_v = 0$, $k^* = 0$)

In summary, Figure 8 shows that increasing $\tau_\theta$ initially reduces surface temperatures while enhancing heat transfer to the deeper parts of the skin. However, as thermal gradients weaken with time, the distinct effects of $\tau_\theta$ fade, and the temperature fields across different cases begin to closely resemble one another.

Figure 9 presents the same simulation cases as Figures 7 and 8, but this time the radial temperature distribution is shown at a fixed depth of $z = 0$ mm, i.e., at the skin surface. As observed in the early stages of the simulation, particularly at 1 s and 5 s, cases with higher values of $\tau_\theta$ exhibit lower temperatures near the center of the laser exposure. This is due to increased effective thermal conductivity, which causes the absorbed heat to spread more quickly from the irradiated zone, resulting in a reduced local temperature rise. This behavior is consistent with what was previously reported in Figures 7 and 8.

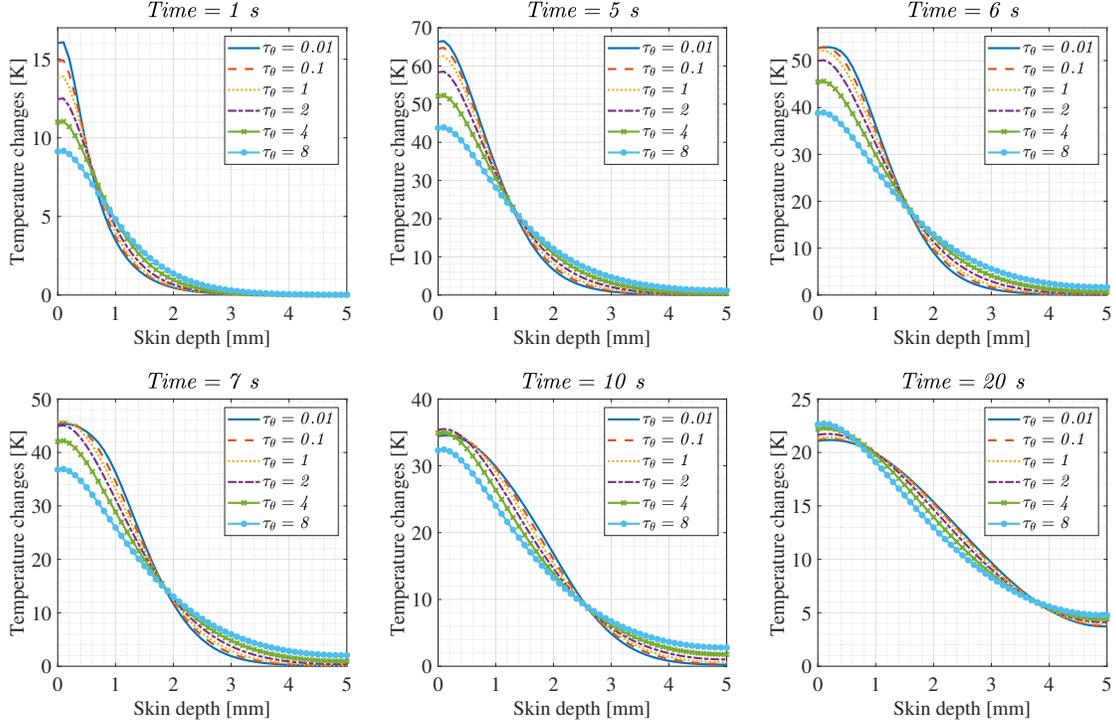

Figure 8: Investigation of temperature profiles in skin tissue depth for different $\tau_\theta$ values at various time steps. ($r = 0$ *mm*, $\tau_q = 1$, $\tau_v = 0$, $k^* = 0$)

As time progresses up to around 10 s the temperature differences between the various $\tau_\theta$ cases diminish. Except for the case with $\tau_\theta = 8$, the other curves converge and exhibit nearly identical temperatures at the surface. This suggests that the effect of $\tau_\theta$ weakens over time as the thermal gradients become less pronounced. However, it is important to note that this convergence occurs only at the surface; as seen in Figure 8, significant differences between the cases still persist at greater depths.

By the final time point at $t = 20$ s, the case with $\tau_\theta = 8$ still maintains a higher temperature near the center compared to the other cases. This trend

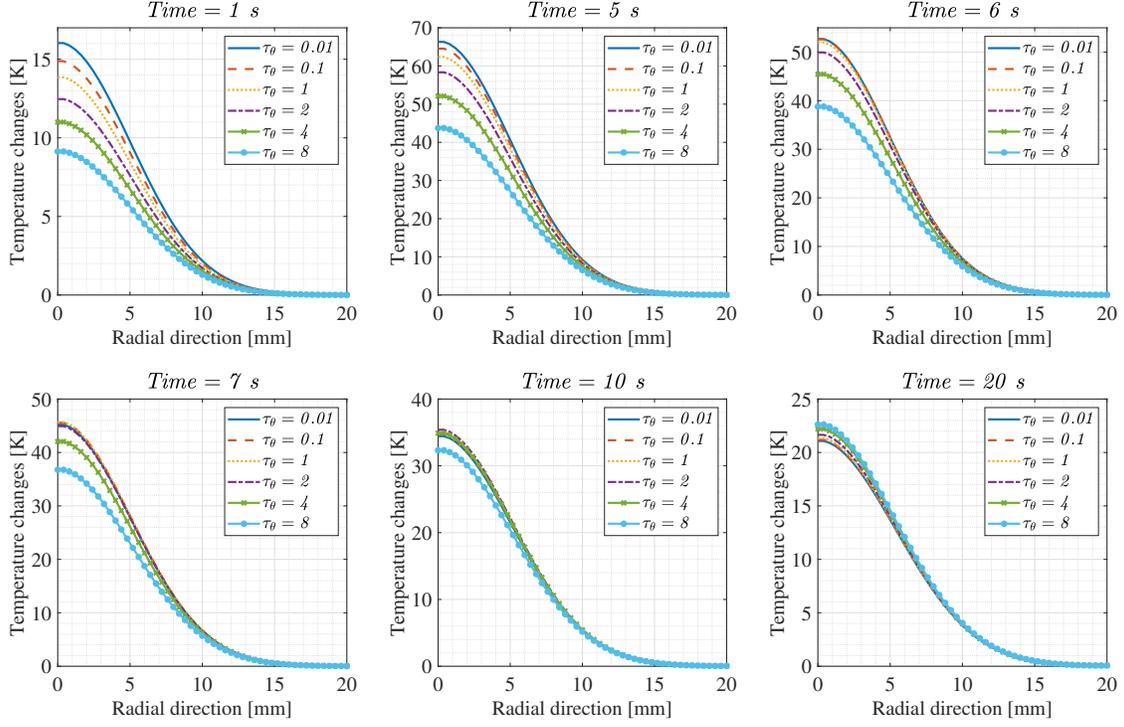

Figure 9: Investigation of temperature profiles in the radial direction of skin tissue for different $\tau_\theta$ values at various time steps. ($z = 0$ mm, $\tau_q = 1$, $\tau_v = 0$, $k^* = 0$)

is consistent with the observations in Figure 7, where at $z = 0$ and $r \approx 3$ mm, the same case also demonstrates the highest temperature at $t = 20$ s. Therefore, Figure 9 reinforces and complements the findings of the previous figures by confirming this thermal behavior in the radial direction.

## 5.3. Analysis of $k^*$

Figure 16 presents a set of simulations distinct from those in Figures 4 through 9. In this figure, the simulations are carried out under the conditions of $\tau_q = 1$ and $\tau_\theta = 1$, while varying the value of $k^*$. This corresponds to a full TPL model where the thermal displacement effects are actively contributing.

A key observation is that during the laser heating phase (i.e., for $t < 5$ s), all temperature responses across different $k^*$ values remain nearly identical. However, after the laser is turned off at $t = 5$ s, the thermal responses begin

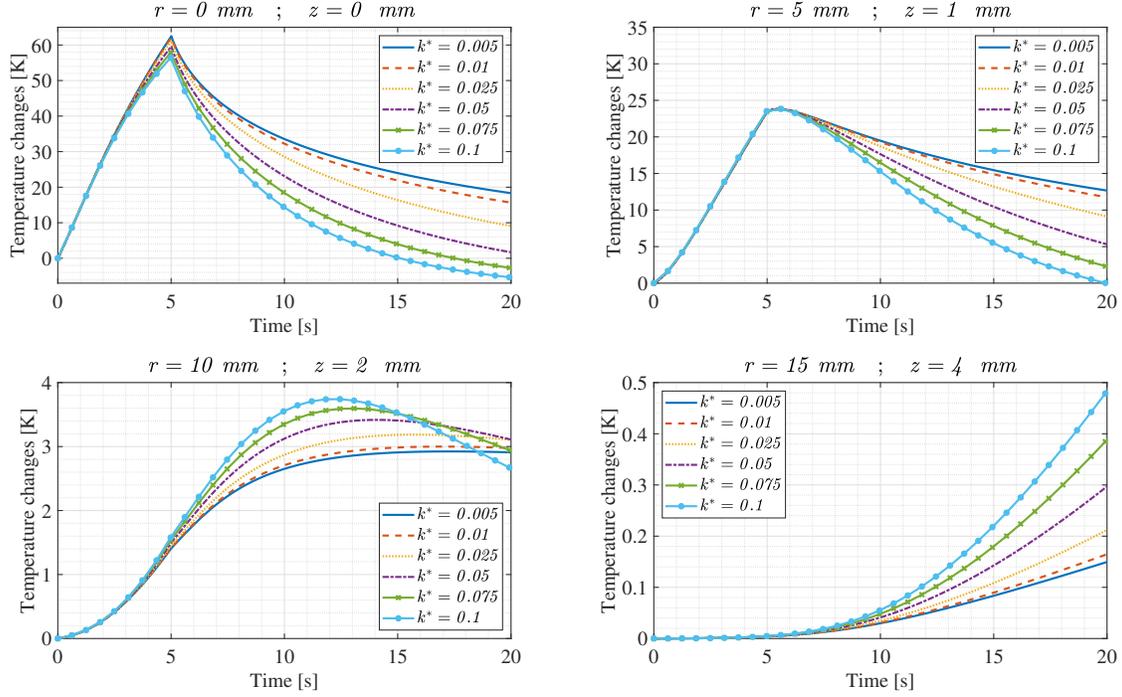

Figure 10: Examining the temperature time history at different locations from the two dimensional modeling platform for different $k^*$ values. ($\tau_q = 1$, $\tau_\theta = 1$, $\tau_v = 0$)

to diverge.

In the first location ($r = 0$ mm, $z = 0$ mm), which corresponds to the laser center, higher values of $k^*$ result in a more rapid post-irradiation temperature decline. This trend is also evident in the second location ($r = 5$ mm, $z = 1$ mm), indicating that larger thermal displacement coefficients enhance the outward propagation of heat.

However, this pattern reverses in the third and fourth locations. For instance, in the third location ($r = 10$ mm, $z = 2$ mm), cases with higher $k^*$ values initially exhibit higher temperatures, but eventually drop below the others, indicating a wave-like behavior. This dynamic temperature evolution suggests delayed energy transfer and possible thermal wavefront movement, which will be further investigated through the spatial profiles in Figures 17 and 18.

Figure 11 validates the wave-like interpretations previously discussed for

Figure 10. This figure illustrates the temperature profiles along the depth direction at the radial location $r = 0$. The results clearly demonstrate that increasing the thermal displacement coefficient $k^*$ induces wave-like thermal behavior in the tissue response.

At $t = 5$ s (top left panel), when the laser is still on, the thermal profiles are nearly identical across all $k^*$ values. However, once the laser is turned off, wave-like behavior emerges for the cases with larger $k^*$ values. From $t = 8$ s to $t = 20$ s, the presence of temperature oscillations (peaks and troughs) becomes more prominent as $k^*$ increases.

These oscillations indicate that larger values of $k^*$ not only amplify the wave-like characteristics of the temperature response but also accelerate the propagation speed of thermal waves within the tissue. Although such behavior is unlikely to be observed in real experimental settings, it remains a subject of theoretical interest and will be further explored and validated through future analyses and experiments.

Figure 18 presents the radial temperature profiles resulting from different values of the thermal displacement coefficient $k^*$. Compared to Figure 17, which shows depthwise profiles, the wave-like behavior observed in this figure is significantly milder along the radial direction.

One major factor contributing to the more prominent wave effects in the depthwise direction is the presence of convective cooling at the skin surface. This cooling rapidly extracts thermal energy from the surface, creating a steep temperature gradient along the $z$-axis that strongly stimulates wave-like responses in the tissue.

Moreover, as depth increases, the energy density absorbed from the laser decreases, further intensifying the temperature gradient in the vertical direction. In contrast, while a Gaussian beam profile is used for the radial direction, the induced temperature gradient is less steep, making the wave-like behavior along the radial axis less pronounced.

Nevertheless, it is important to note that the wave response in the radial direction can be enhanced by adjusting the laser spot size. In later sections, this approach will be employed as part of a strategy to determine suitable $k^*$ values based on controlled stimulation of radial thermal wave patterns.

*5.4. Analysis of $\tau_v$*

Figure 13 explores the thermal behavior under varying values of the thermal displacement phase lag $\tau_v$, while keeping other parameters fixed: $\tau_q = 1$,

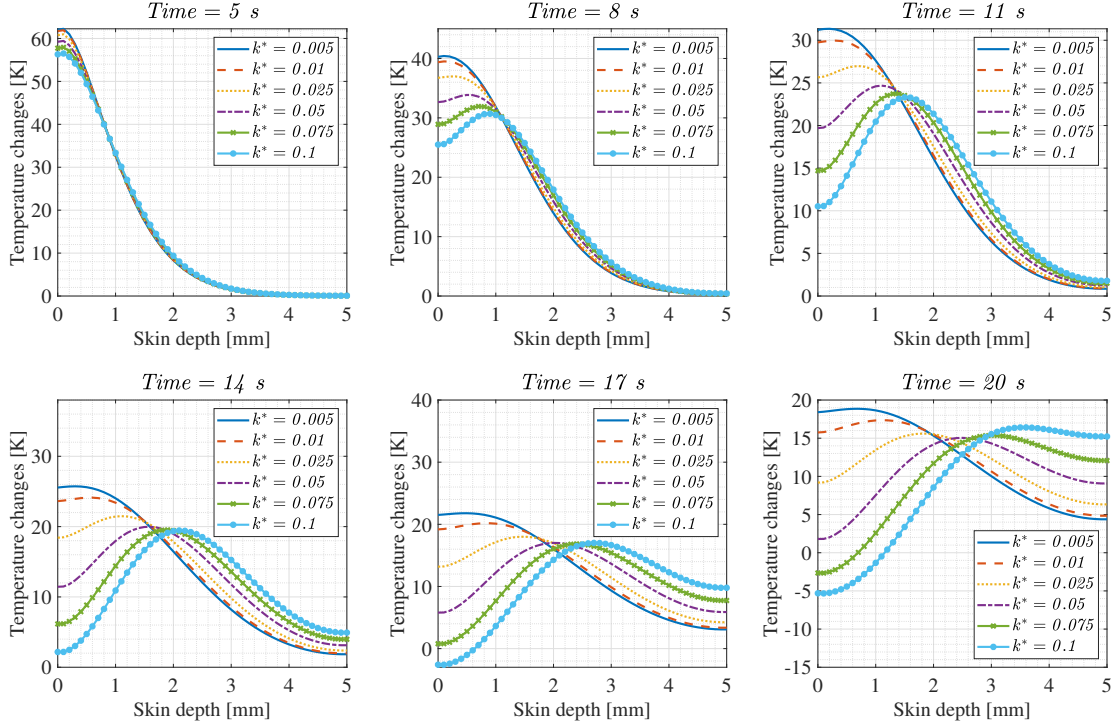

Figure 11: Investigation of temperature profiles in skin tissue depth for different $k^*$ values at various time steps. ($r = 0$, $\tau_q = 1$, $\tau_\theta = 1$, $\tau_v = 0$)

$\tau_\theta = 1$, and $k^* = 0.1$. Unlike previous simulations where $\tau_v$ was set to zero, in this case multiple values are assigned to investigate its independent influence.

As observed in the temperature-time curves, the influence of $\tau_v$ closely resembles that of $\tau_\theta$. This similarity can be confirmed by comparing Figure 13 with Figure 7. Moreover, a partial explanation lies in the structure of the governing equation (e.g., Equation (4)), where both $\tau_v$ and $\tau_\theta$ contribute to similar gradient-based terms.

Despite the observed similarity, minor differences between the effects of $\tau_v$ and $\tau_\theta$ are also evident. To evaluate these distinctions more precisely, additional profiles shown in Figures 14 and 15 must be examined in the following analysis.

Figure 14 further develops the comparison between the influence of $\tau_\theta$ and $\tau_v$ by examining their effects on the depthwise temperature distribution.

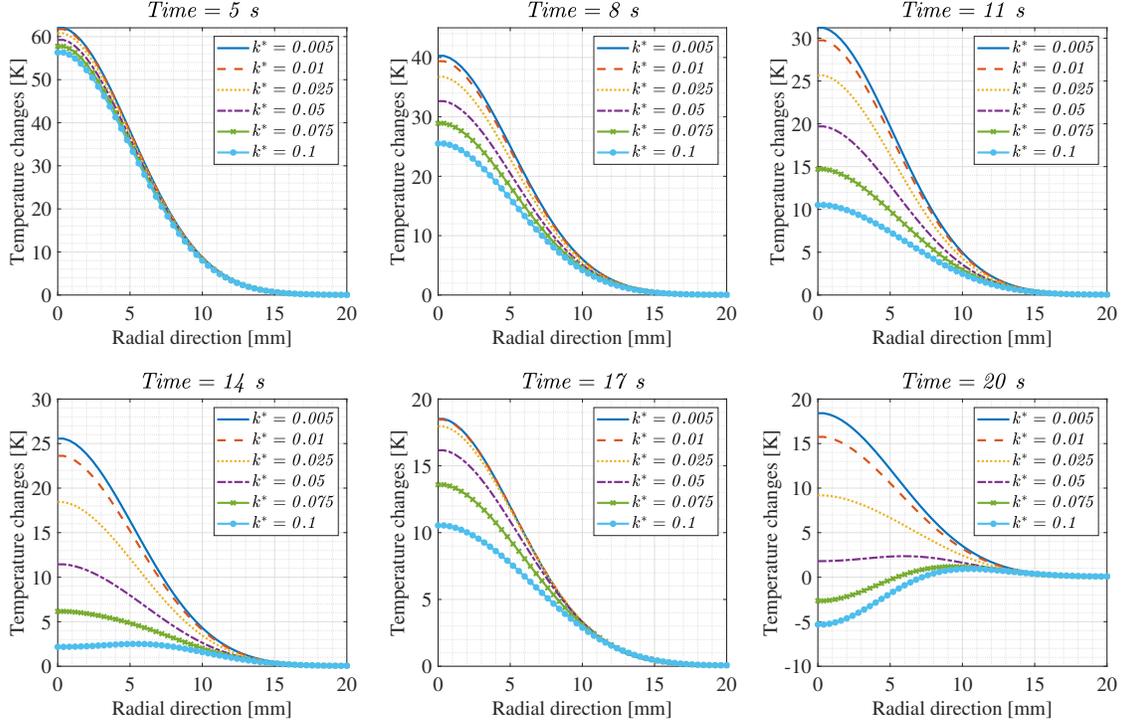

Figure 12: Investigation of temperature profiles in the radial direction of skin tissue for different $k^*$ values at various time steps. ($z = 0$, $\tau_q = 1$, $\tau_\theta = 1$, $\tau_v = 0$)

While similarities exist between the two, this figure reveals a key distinction. As previously observed in Figure 8, increasing $\tau_\theta$ raised the thermal conductivity during heating phases and lowered it during cooling phases. In contrast, $\tau_v$ consistently enhances thermal conductivity, regardless of whether the tissue is heating or cooling.

As clearly illustrated across all six time snapshots in Figure 14, higher values of $\tau_v$ result in stronger thermal diffusion throughout the tissue depth. This effect is not limited to the heating stage (before $t = 5\,\text{s}$), but continues during the cooling phase as well, causing thermal energy to spread more uniformly through the domain.

Moreover, the increased conductivity associated with larger $\tau_v$ values reduces the thermal wave amplitudes induced by the $k^*$-driven wave behavior. It is important to note that these simulations are conducted with $k^* = 0.1$,

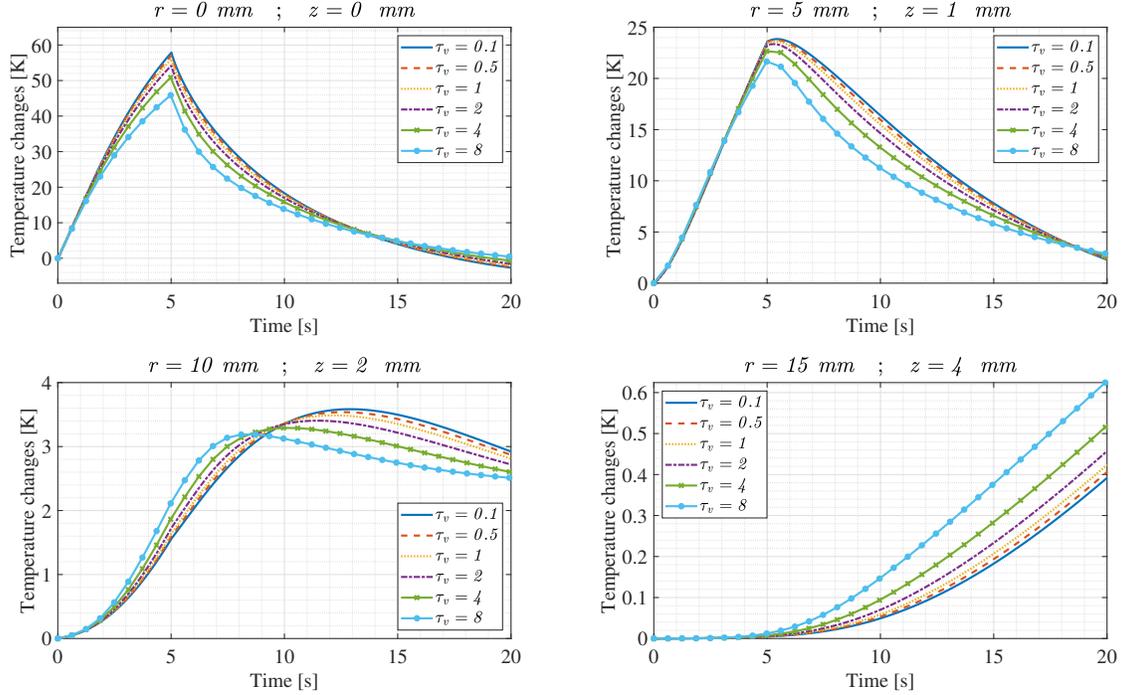

Figure 13: Examining the temperature time history at different locations from the two dimensional modeling platform for different $\tau_v$ values. ($\tau_q = 1$, $\tau_\theta = 1$, $k^* = 0.1$)

which independently induces wave-like patterns in the temperature field, as previously demonstrated in Figure 11. Therefore, while $\tau_v$ does not directly cause wave phenomena, it clearly modulates them by enhancing diffusion and smoothing out oscillatory thermal features.

Figure 15 shows that, similar to the depthwise results in Figure 14, increasing the thermal displacement phase lag $\tau_v$ reduces the amplitude of the thermal wave patterns induced by the non-dimensional conductivity parameter $k^*$. A particularly interesting behavior emerges when observing how the temperature profiles evolve over time.

Up to $t = 14$ s, the case with the largest $\tau_v$ value ($\tau_v = 8$) consistently shows lower temperatures near the center of the domain. This is attributed to the higher thermal conductivity effect introduced by a large $\tau_v$, which promotes more rapid heat spreading. However, starting from $t = 15$ s and continuing through $t = 20$ s, the temperature of the same case becomes higher

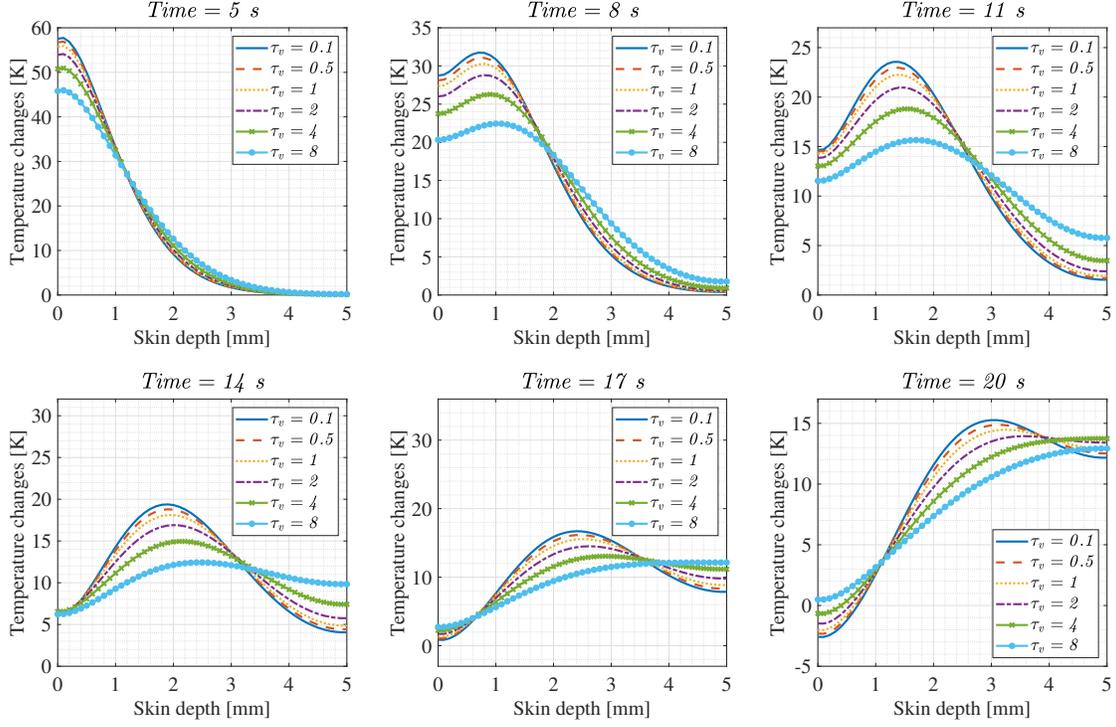

Figure 14: Investigation of temperature profiles in skin tissue depth for different $\tau_v$ values at various time steps. ($r = 0$, $\tau_q = 1$, $\tau_\theta = 1$, $k^* = 0.1$)

than the others.

This apparent reversal is due to the wave-like behavior of the system. At later stages, the lower $\tau_v$ cases exhibit more pronounced thermal oscillations, particularly in the central region ($r = 0$), causing their temperatures to dip due to thermal wavefront troughs. In contrast, the large $\tau_v$ case suppresses these waveforms through enhanced heat diffusion, resulting in smoother and more stable temperature profiles.

In essence, while a large $\tau_v$ continues to promote thermal diffusion, it also acts to dampen the instability caused by thermal waves. As a result, it helps maintain a relatively elevated temperature in regions where wave-induced dips are prominent in low-$\tau_v$ cases.

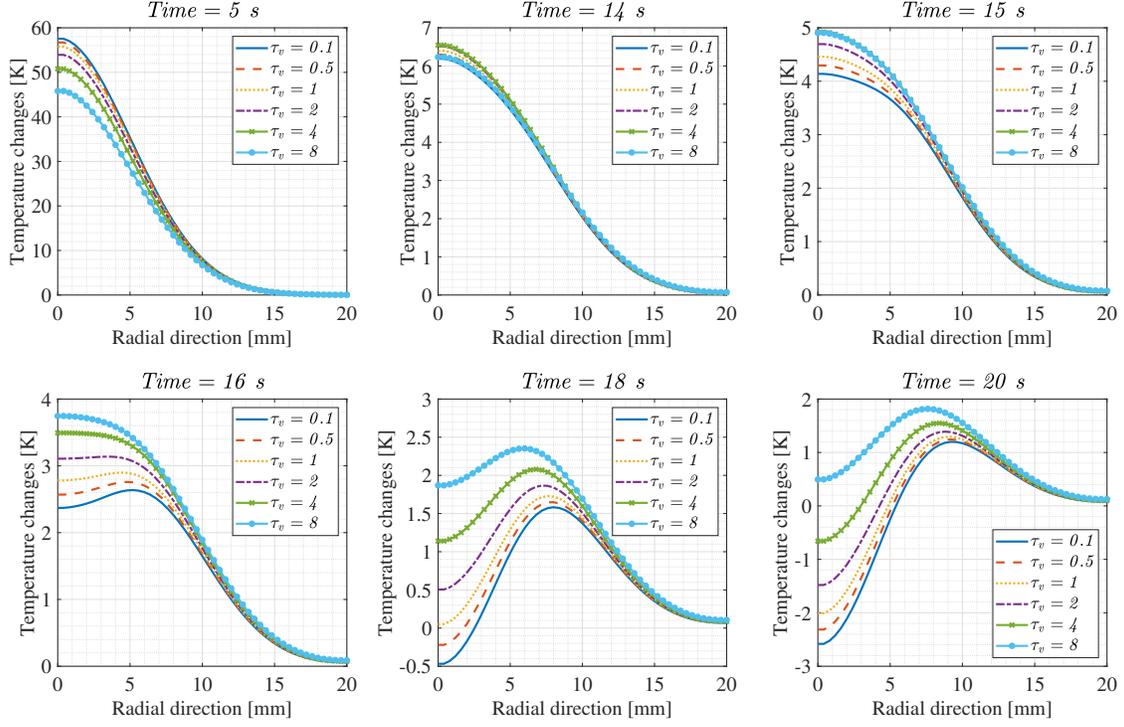

Figure 15: Investigation of temperature profiles in the radial direction of skin tissue for different $\tau_v$ values at various time steps. ($z = 0$, $\tau_q = 1$, $\tau_\theta = 1$, $k^* = 0.1$)

### 5.5. Analysis of different thermal conductivity measurements

As mentioned in the section Experimental Setup and Simulation Parameters, the thermal conductivity used in the simulations may not accurately reflect the actual value for biological skin tissue due to experimental uncertainty or modeling assumptions. Additionally, as previously observed in the analysis of other parameters such as $\tau_\theta$ and $\tau_v$, similar effects to those caused by varying the thermal conductivity $k$ can occur. Hence, Figures 16, 17, and 18 present the temperature response of the skin for different thermal conductivity values to support parameter evaluation and comparative studies.

Figure 16 clearly shows that increasing $k$ leads to a lower temperature in regions directly heated by the laser (e.g., $r = 0$, $z = 0$) due to faster heat conduction away from the center. Conversely, regions that warm up

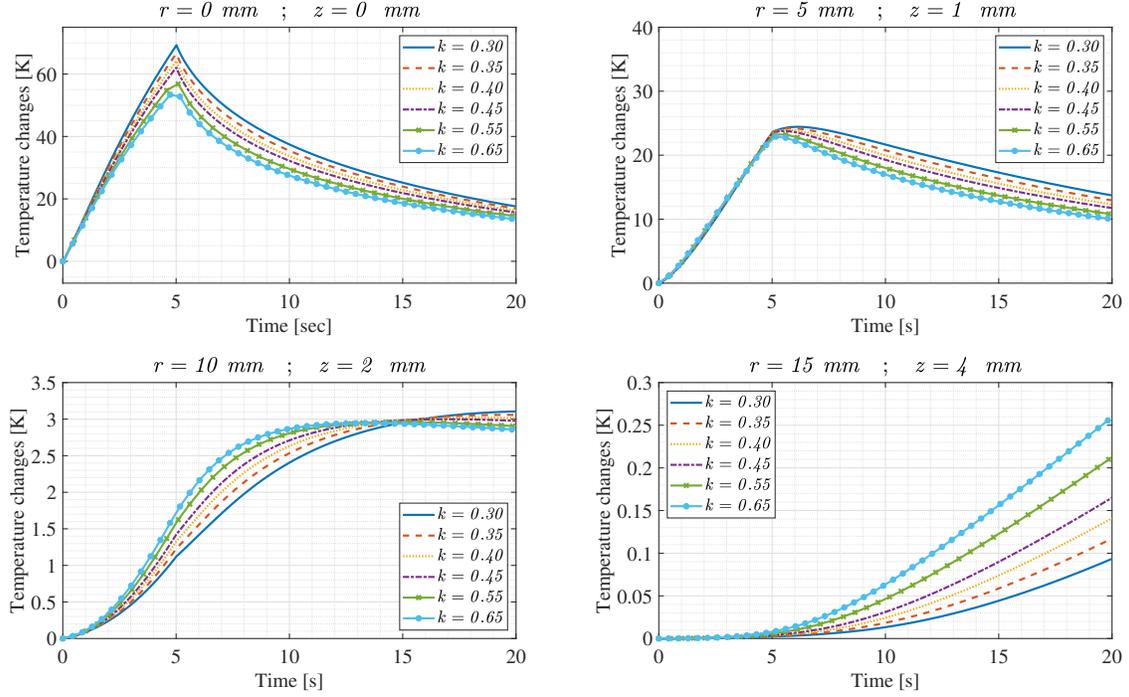

Figure 16: Examining the temperature time history at different locations from the two dimensional modeling platform for different thermal conductivity values. ($\tau_q = 1$, $\tau_\theta = 1$, $\tau_v = 0.1$, $k^* = 0.01$)

only through thermal diffusion such as $r = 15$ mm and $z = 4$ mm experience higher temperatures as $k$ increases, since more heat is transferred into these zones over time.

This trend is also confirmed in Figure 17, which displays temperature profiles along the skin depth direction. While temperature differences are initially small at $t = 5$ s, they become more pronounced over time. By $t = 20$ s, larger $k$ values result in greater temperature increases at deeper parts of the skin, confirming enhanced thermal propagation with higher conductivity.

Similarly, Figure 18 illustrates the temperature profiles along the radial direction. Here too, increasing $k$ broadens the temperature distribution outward from the center. Although the effect is less steep than in the depth direction due to the Gaussian laser profile, the overall trend persists: higher $k$ values cause heat to spread more efficiently in all directions.

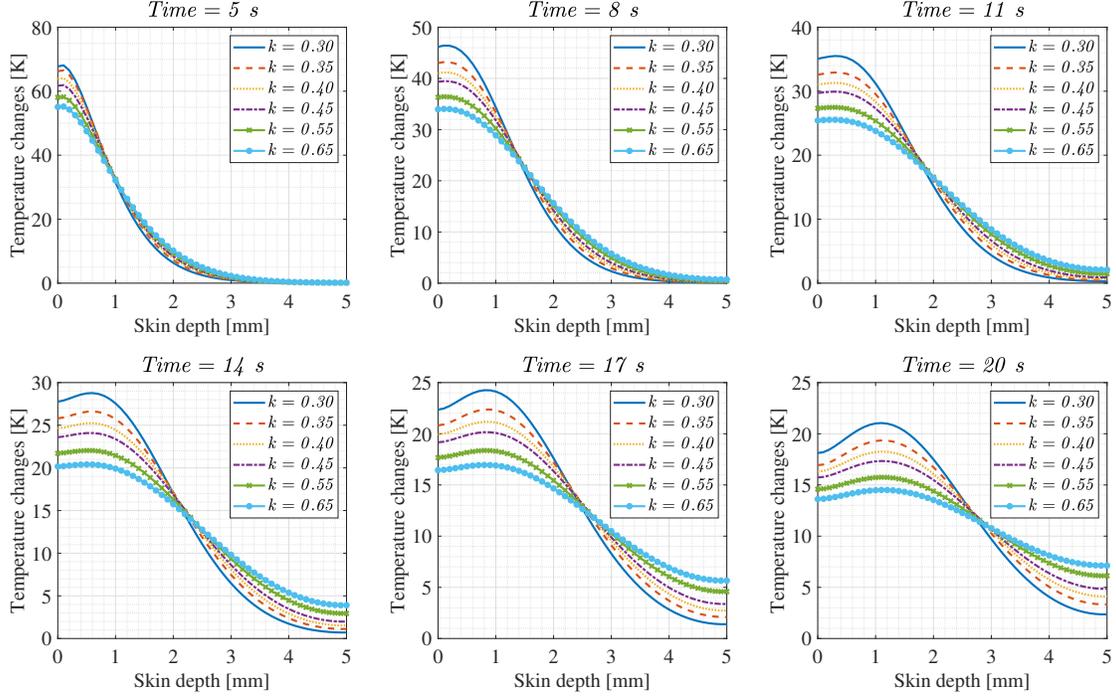

Figure 17: Investigation of temperature profiles in skin tissue depth for different thermal conductivity values at various time steps. ($r = 0$, $\tau_q = 1$, $\tau_\theta = 1$, $\tau_v = 0.1$, $k^* = 0.01$)

In conclusion, these figures confirm that thermal conductivity $k$ significantly influences the spatio-temporal thermal response of the tissue. Understanding this behavior is critical when designing experiments to estimate unknown phase lag parameters, as it helps to decouple the effects of $k$ from other thermal mechanisms.

## 6. Design of experiment and results

The experimental design for determining the four investigated parameters entails several inherent challenges, such as sensor misalignment relative to the intended measurement locations, uncertainties in the input thermal energy due to non-ideal efficiency of the laser source, and limitations in sensor capability to record point-specific temperatures. Furthermore, uncertainties in thermal properties such as thermal conductivity ($k$) and specific heat

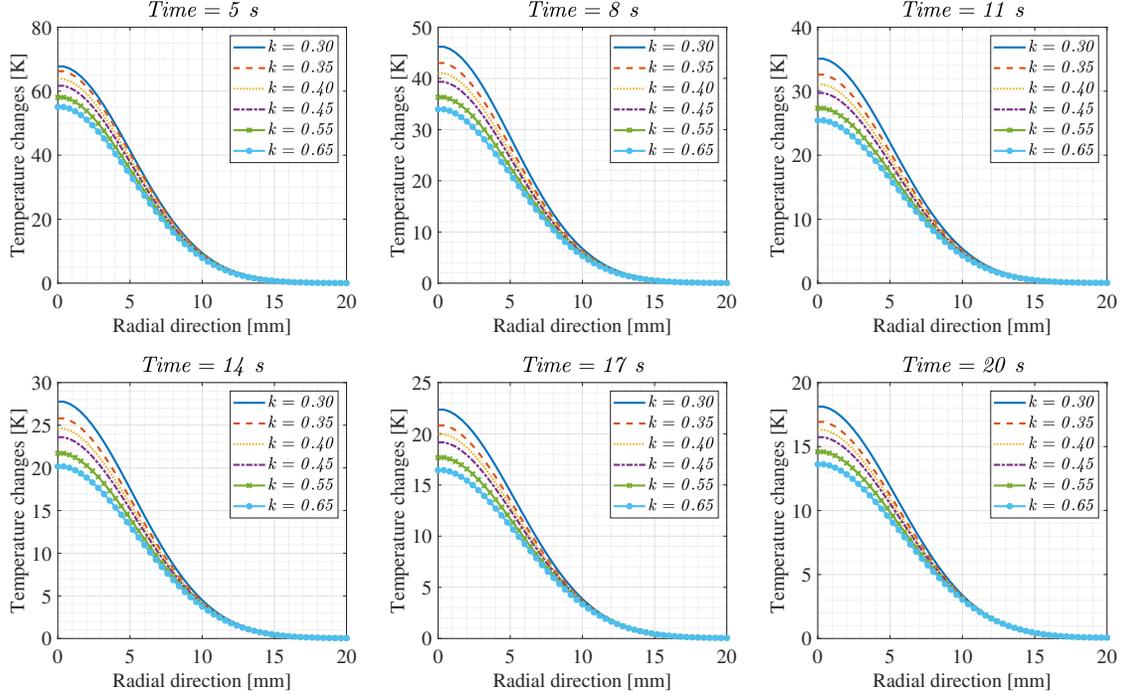

Figure 18: Investigation of temperature profiles in the radial direction of skin tissue for different thermal conductivity values at various time steps. ($z = 0$, $\tau_q = 1$, $\tau_\theta = 1$, $\tau_v = 0.1$, $k^* = 0.01$)

capacity introduce additional experimental complexity. As a result, the experimental setup must be carefully designed to mitigate the influence of these uncertainties.

As illustrated in simulations presented in Figures 3 to 18, certain parameters produce thermally similar effects or responses, potentially leading to overlapping influences on the measured data. Hence, it is essential to ensure that the effects of individual parameters are decoupled within the experimental design. For instance, Figures 7 through 9, when compared with Figures 16 through 18, reveal notable similarities between the effects of the temperature gradient phase lag $\tau_\theta$ and the thermal conductivity $k$. Given the lack of definitive experimental studies on the thermal properties of the selected skin tissue, the experimental design must ensure that uncertainties in thermal conductivity do not propagate into errors in estimating $\tau_\theta$.

A similar concern is observed in the comparison of Figures 10 to 12 with Figures 13 to 15, where the behavior of the thermal displacement phase lag $\tau_v$ closely resembles that of an increased thermal conductivity. This phenomenon is also supported analytically by Equation (13), which shows that the presence of $\tau_v$ reduces the role of $k^*$ in generating thermal wave effects. Therefore, to accurately identify each parameter, distinct experimental protocols must be developed to minimize the influence of other parameters and reduce potential experimental errors.

$$\begin{aligned}
k &\left[\left(\frac{1}{r}\frac{\partial \theta}{\partial r} + \frac{\partial^2 \theta}{\partial r^2} + \frac{\partial^2 \theta}{\partial z^2}\right) + \tau_\theta \frac{\partial}{\partial t}\left(\frac{1}{r}\frac{\partial \theta}{\partial r} + \frac{\partial^2 \theta}{\partial r^2} + \frac{\partial^2 \theta}{\partial z^2}\right)\right] \\
+ k^* &\left[\left(\frac{1}{r}\frac{\partial}{\partial r}\left(\int_0^t \theta d\tau\right) + \frac{\partial^2}{\partial r^2}\left(\int_0^t \theta d\tau\right) + \frac{\partial^2}{\partial z^2}\left(\int_0^t \theta d\tau\right)\right)\right] \\
&+ k^* \left[\tau_v \left(\frac{1}{r}\frac{\partial \theta}{\partial r} + \frac{\partial^2 \theta}{\partial r^2} + \frac{\partial^2 \theta}{\partial z^2}\right)\right] \\
+ &\left[Q_L(r,z,t) - \rho c \frac{\partial \theta}{\partial t}\right] + \tau_q \left[\frac{\partial Q_L(r,z,t)}{\partial t} - \rho c \frac{\partial^2 \theta}{\partial t^2}\right] = 0
\end{aligned} \quad (13)$$

In the next phase, a series of experiments will be designed to identify and evaluate the unknown thermal parameters. These tests will be conducted using the experimental setup and laser exposure on skin tissue, as illustrated in Figure 19. The results obtained from these experiments will be directly compared with simulations replicating the same conditions.

It is important to note that a laser beam with a diameter of 3 mm was used in these experiments. Furthermore, the temperatures recorded by the sensors do not represent the exact point-wise temperature at a focused spot. Due to the Field Of View (FOV) of the sensors, the measured values correspond to the average temperature across an arc-shaped area in front of each sensor. This characteristic has also been accounted for and modeled in the simulation domain to ensure accurate comparison.

To ensure consistency and reliability of the experimental data, each test was repeated between 10 and 20 times. This repetition was intended to confirm the repeatability of the thermal responses under identical conditions.

*6.1. Derivation $\tau_q$*

Based on the analysis of Figures 4,b5, and 6, as well as Equation (13), it is evident that the influence of $\tau_q$ on the thermal response of skin tissue is primarily dependent on the rate of laser irradiation, i.e., the time derivative of the laser heat source function $Q_L$. As the value of $\tau_q$ increases, the pulsed laser function acts more like an impulse, causing a faster rise in temperature and resulting in higher thermal peaks during the irradiation period. However, this effect is restricted to the duration of laser exposure; once the irradiation ceases, the sudden drop in the laser pulse similarly behaves like an impulse in the cooling direction.

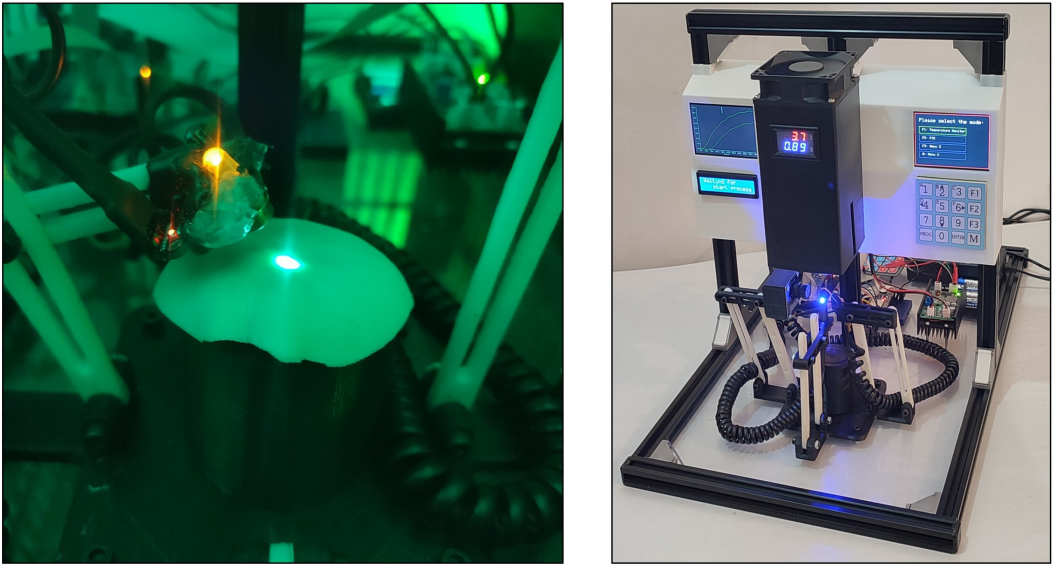

Figure 19: The experimental system employed for conducting 450 nm laser irradiation tests on skin tissue specimens.

To estimate $\tau_q$, an experimental protocol must be designed where all test conditions remain identical except for the time derivative of $Q_L$, or equivalently, the rate at which energy is delivered to the tissue. By examining the variation in thermal response resulting from changes in the derivative of $Q_L$, an estimation of $\tau_q$ can be made.

In the proposed setup, two types of laser irradiation with the same total energy but different rates of energy delivery are applied to the skin tissue. The temperature is recorded by Sensor 1. The recorded temperatures under the two conditions are expected to exhibit differences due to the variation in heating rates. These two cases, which differ solely in the temporal rate of $Q_L$, are then simulated under multiple values of $\tau_q$.

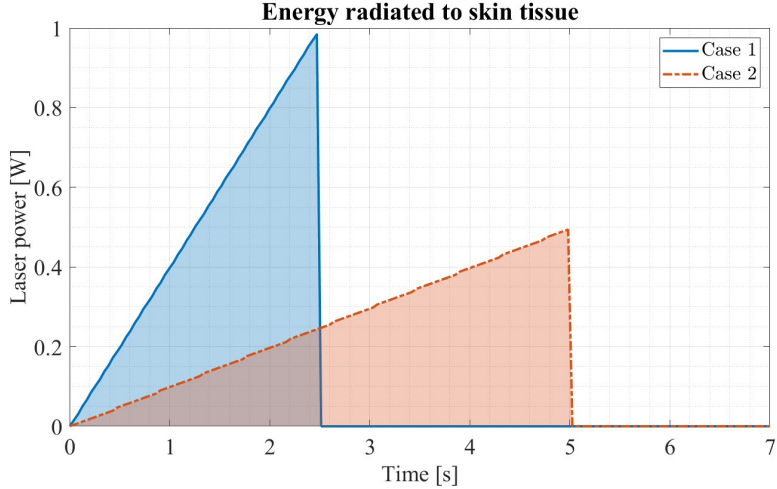

Figure 20: Irradiation energy profiles corresponding to Cases 1 and 2, as recorded from the microcontroller within the experimental setup for laser-tissue interaction.

Simulation results are expected to show that smaller values of $\tau_q$ correspond to smaller temperature differences between the two cases, whereas larger values of $\tau_q$ result in increased differences. By comparing the experimental differences with those obtained from the simulations for various $\tau_q$ values, a plausible range for $\tau_q$ can be determined.

To implement the previously described conditions for evaluating and estimating the value of $\tau_q$, two irradiation cases (Cases 1 and 2), as presented in Figure 20, were designed such that the temporal slope of Case 1 was twice that of Case 2. These cases were programmed into the microcontroller of the experimental setup and were executed repeatedly on skin tissue samples.

In parallel, simulations were also conducted for the same two cases under various values of $\tau_q$, with the results shown in Figure 23. It is worth noting that while the experimental temperature responses shown in Figure 22 resemble some of the simulated outcomes in Figure 23, the comparison methodology is not based solely on overall waveform similarity. Instead, the approach is to identify which value of $\tau_q$ produces a maximum temperature difference between the two cases that best matches the experimentally measured maximum temperature difference.

As observed, the simulation corresponding to $\tau_q = 0.1$ yields results most consistent with the experimental data. For values of $\tau_q$ below 0.1, reliable interpretation is not feasible using the current experimental setup, as the temporal resolution of the infrared sensors is limited to 0.1 seconds. More accurate experimental systems would be required for validating $\tau_q$ values in that lower range.

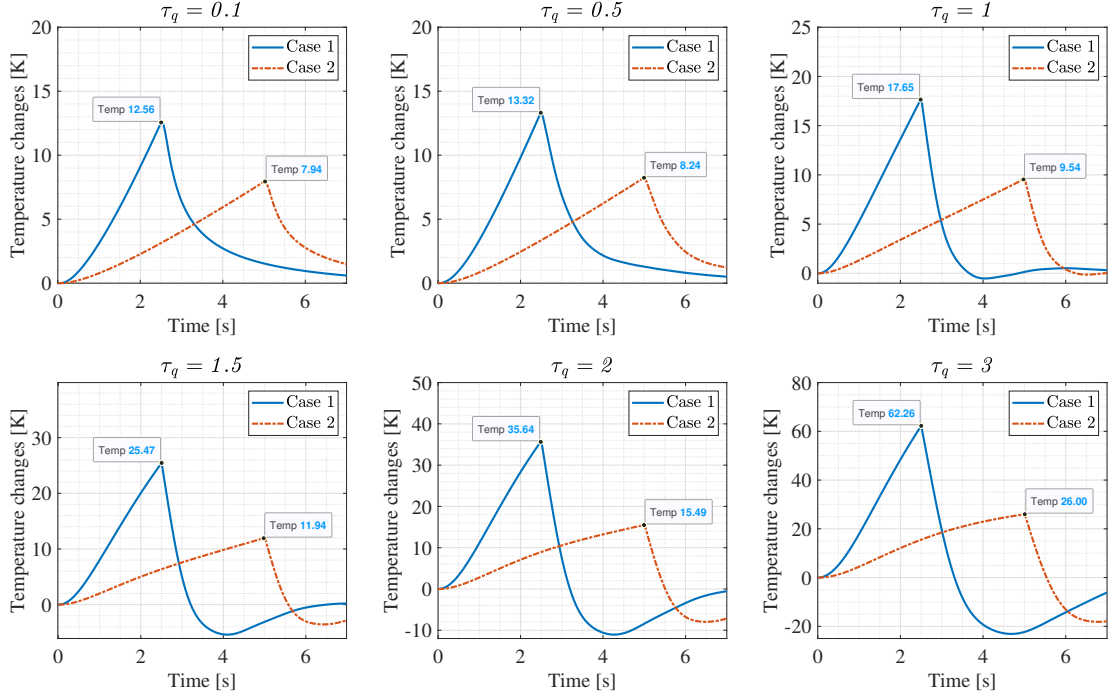

Figure 21: Simulated thermal responses for Cases 1 and 2 (with energy profiles shown in Figure 20), under varying values of the heat flux phase lag $\tau_q$.

*6.2. Derivation $\tau_\theta$*

According to the conducted simulations, the estimation of $\tau_\theta$ appears to be more challenging than that of the other parameters. One primary reason for this difficulty is the considerable similarity between the influence of $\tau_\theta$ and that of variations in thermal conductivity. As a result, any error in the assumed value of thermal conductivity used in the simulations particularly if it deviates significantly from the actual thermal properties of the tissue can interfere with the accurate determination of $\tau_\theta$.

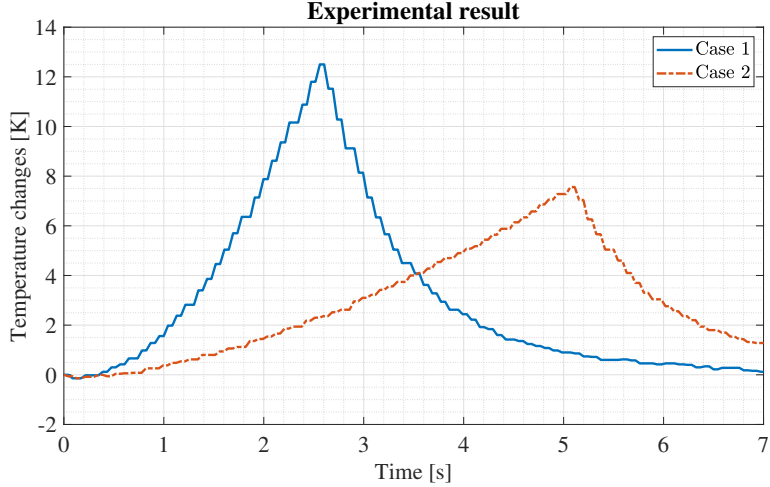

Figure 22: Experimentally measured temperature responses of skin tissue under laser exposure for Cases 1 and 2.

This challenge is evident when comparing Figures 7 to 9 with Figures 16 to 18, where the thermal responses induced by $\tau_\theta$ closely resemble those caused by changes in thermal conductivity. Another reason for the difficulty lies in the relatively weak influence of $\tau_\theta$ on the thermal responses compared to other parameters.

To mitigate such ambiguities, emphasis must be placed on the unique effect of $\tau_\theta$. Detailed examination of the simulation results reveals that during heating phases, when the local temperature increases, $\tau_\theta$ acts to effectively enhance the thermal conductivity. Conversely, during cooling phases when temperature decreases it tends to reduce the effective thermal conductivity. This phenomenon is also supported by the structure of Equation (13).

The practical manifestation of this behavior is a difference in the slope of the heating and cooling segments of the temperature-time curves, which can be captured by the temperature sensors particularly Sensor 1. Based on this observation, a method is proposed wherein $\tau_\theta$ is estimated by applying periodic pulsed laser irradiation with varying periods, and then examining the symmetry or asymmetry in the slopes of the heating and cooling phases.

*6.3. Derivation $k^*$*

The estimation of $k^*$ is expected to be more straightforward compared to the other parameters, as its influence on the thermal response is distinct and can be easily distinguished from the effects of other parameters. According to Equation (13), the inclusion of $k^*$ introduces a temporal integral term that results in a wave-like behavior in the thermal response. This phenomenon is clearly illustrated in Figures 11 and 12.

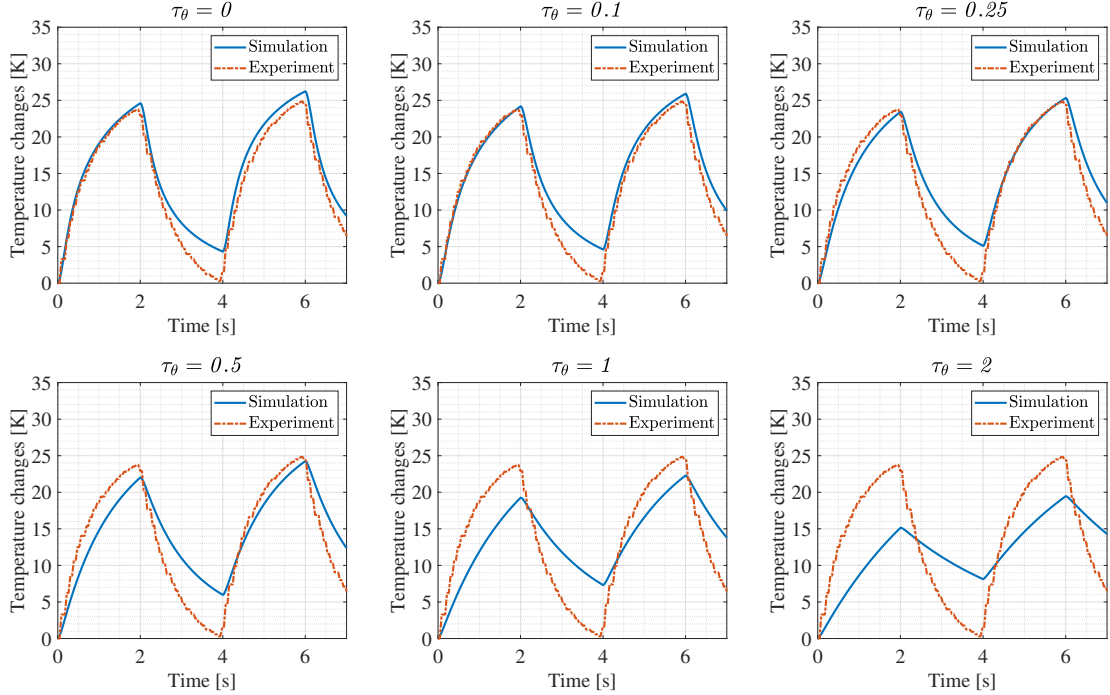

Figure 23: Simulated temperature responses under periodic pulsed laser irradiation for various values of $\tau_\theta$, compared against corresponding experimental results.

As observed in Figure 12, the wave-like nature of heat transfer induced by $k^*$ can be readily investigated using Sensors 1, 2, and 3. These measurements enable the identification of oscillatory thermal patterns in the skin tissue, thus facilitating the determination of the $k^*$ parameter with high confidence.

To evaluate the effect of the thermal displacement coefficient $k^*$, a 3-second laser pulse was applied, and the temperature response was recorded up to $t = 30$ s using Sensors 1 and 2. As shown in Figure 24, simulations were conducted for $k^*$ values of 0, 0.001, 0.01, and 0.1. The results demonstrate that increasing $k^*$ leads to the emergence of wave-like thermal behavior: after the laser is turned off, the temperature recorded at the center of irradiation by Sensor 1 becomes lower than that of Sensor 2. This reversal in relative temperatures is a hallmark of thermal wave propagation, as previously discussed. It should be noted that the "sensors" in this context refer to virtual sensors, which capture temperature data from specific spatial regions using a FOV model. This modeling approach ensures consistency with experimental sensor measurements, which are presented in Figure 25.

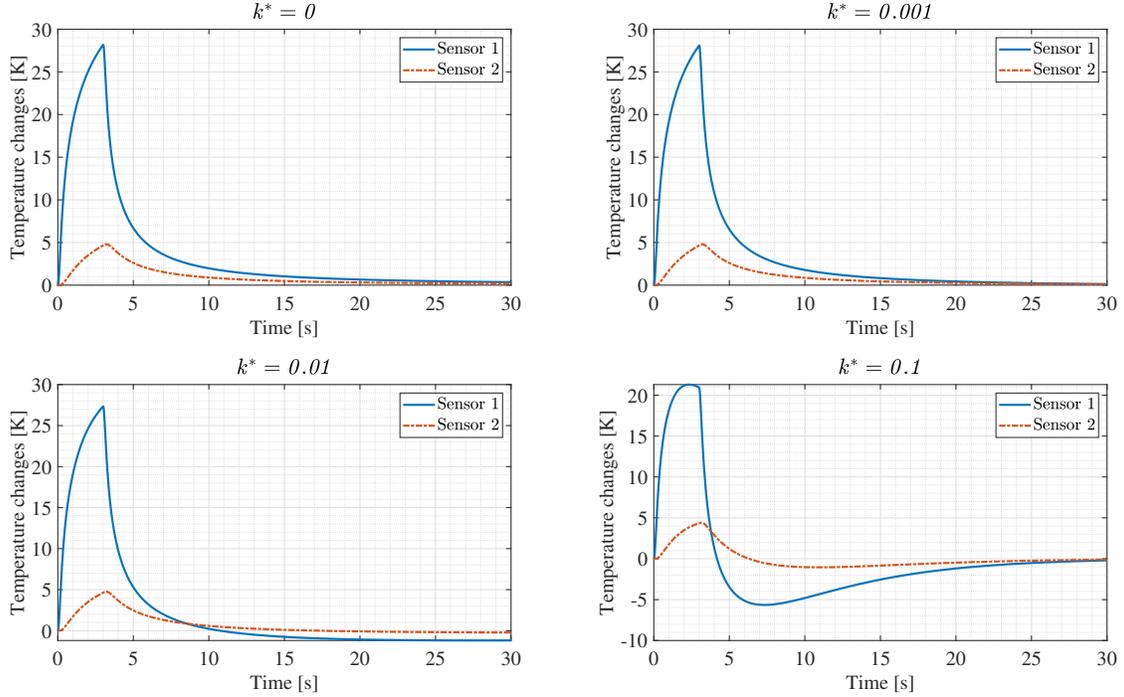

Figure 24: Simulated thermal responses to 3-second laser pulses with different $k^*$ values, used for assessing the appropriate range of the thermal displacement coefficient.

By comparing Figures 24 and 25, it can be concluded that within the scale and resolution of the current study, the value of $k^*$ can be considered negligible. In all experimental trials, the temperature recorded by Sensor 1 was never observed to fall below that of Sensor 2. This indicates an absence of thermal wave behavior in the examined scale and suggests that the contribution of $k^*$ to the heat transfer dynamics in this setup is minimal.

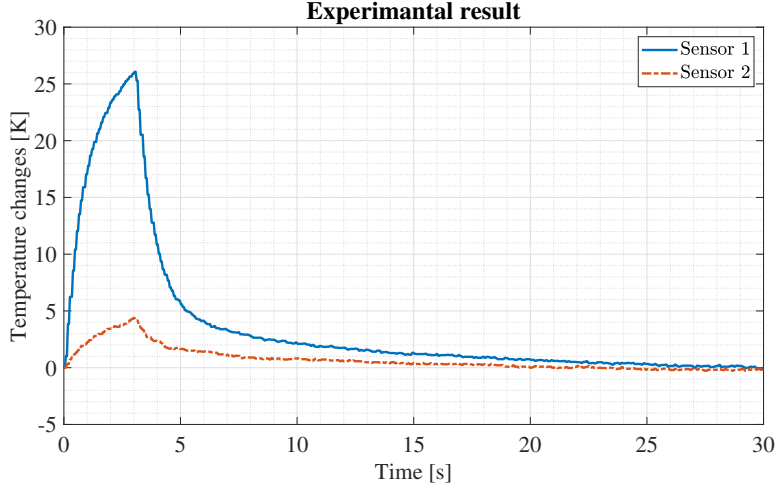

Figure 25: Temperature measurements obtained from Sensors 1 and 2 during a 3-second laser exposure, utilized for the estimation of the thermal displacement coefficient $k^*$.

*6.4. Derivation $\tau_v$*

By analyzing Equation (13), it can be observed that the term

$$k^*\tau_v \left( \frac{1}{r}\frac{\partial \theta}{\partial r} + \frac{\partial^2 \theta}{\partial r^2} + \frac{\partial^2 \theta}{\partial z^2} \right), \tag{14}$$

behaves analogously to the thermal conductivity term

$$k \left( \frac{1}{r}\frac{\partial \theta}{\partial r} + \frac{\partial^2 \theta}{\partial r^2} + \frac{\partial^2 \theta}{\partial z^2} \right). \tag{15}$$

Furthermore, a comparison between Figures 13 to 15 and Figures 16 to 18 reveals that the influence of $\tau_v$ on the thermal response is virtually identical to the effect of increasing thermal conductivity. Under such circumstances, the presence of $\tau_v$ does not introduce a distinct physical meaning; instead, the term $k^*\tau_v$ effectively adds to the overall thermal conductivity of the material.

In other words, if $k^*\tau_v$ is non-zero, the experimentally measured thermal conductivity of any material inherently includes the contribution of this term, thereby rendering the separate definition of $\tau_v$ redundant. Therefore, it can be concluded that the influence of $\tau_v$ may be disregarded in such scenarios.

## 7. Synthesis and Discussion

In recent years, a substantial number of studies have been conducted in the field of heat transfer modeling in skin tissue and other similar biological materials using SPL, DPL, and TPL approaches. Some of these studies have employed unusually large values for the phase lag parameters, often citing relatively older references as their basis. For instance, several papers have reported phase lag values as high as 15 *s*, which raises concerns about accuracy and applicability in light of recent advances in microcontroller technologies and thermal sensing systems. Furthermore, based on more recent research, assigning such large values (e.g., 16 to 32 *s*) to the phase lag parameters may also render the system thermodynamically ill-posed [53, 54]. Additionally, there are studies that, in contrast to those using large phase lag values, have employed much smaller values, typically in the range of less than 0.5 *s* seconds for the phase lag parameters [11, 55–64].

As discussed in this section, the four principal phase lag parameters have been examined and evaluated in detail. Table 3 presents these parameters within a proposed range, based on the current study. However, it should be emphasized that more accurate identification of these values requires further investigation supported by high-resolution sensors and advanced experimental platforms in future studies.

Table 3: Estimated values or practical assumptions for the phase lag parameters used in this study

| Parameter | Symbol | Estimated Value |
|---|---|---|
| Heat Flux Phase Lag (s) | $\tau_q$ | $\tau_q < 0.5$ (approximate range) |
| Temperature Gradient Phase Lag (s) | $\tau_\theta$ | $\tau_\theta < 0.25$ (approximate range) |
| Thermal Displacement Coefficient | $k^*$ | $\approx 0$ (negligible) |
| Thermal Displacement Phase Lag (s) | $\tau_v$ | 0 (assumed value) |

According to the authors, offering updated and constrained ranges for phase lag values can be of significant importance, especially given the wide variation reported in the literature. This highlights a clear need for more pre-cise experimental validation, which this study aims to contribute to. Moreover, the current work attempts to isolate the individual effects of each phase lag parameter so that the estimation of one is not confounded by the uncertainty or error in others.

## 8. Conclusion

Experimental validation and parameter extraction techniques play a major role in the successful development of bioheat transfer mechanisms in complex biological tissues, like skin. This study thoroughly investigated the influence of the four primary parameters within the Three-Phase Lag (TPL) heat transfer model: the heat flux phase lag ($\tau_q$), temperature gradient phase lag ($\tau_\theta$), thermal displacement coefficient ($k^*$), and thermal displacement phase lag ($\tau_v$). A systematic experimental and numerical analysis was conducted, leading to the following key observations:

- Increasing $\tau_q$ significantly intensifies both heating and cooling rates, potentially causing unrealistic thermal instabilities at higher values.

- The temperature gradient phase lag, $\tau_\theta$, was found to enhance heat diffusion during heating phases and reduce it during cooling phases, demonstrating a subtle but measurable influence.

- The thermal displacement coefficient, $k^*$, was associated with distinct wave-like thermal responses; however, experimental validation indicated that its effect is negligible at the investigated scales.

- The thermal displacement phase lag, $\tau_v$, primarily acts as an additional effective thermal conductivity term, showing no unique physical distinction from standard conductivity effects.

By comparing simulation results with carefully designed bovine skin experiments, realistic ranges for these parameters were identified:

- $\tau_q < 0.5$ s

- $\tau_\theta < 0.25$ s

- $k^* \approx 0$

- $\tau_v = 0$

These identified parameter ranges differ notably from the large values reported in previous literature, emphasizing the importance of precise experimental validation and parameter isolation. The findings significantly improve the accuracy of bioheat transfer modeling, providing a robust basis for future research and clinical thermal therapy applications.

**Acknowledgments**

TR thank the Deutsche Forschungsgemeinschaft (DFG, German Research Foundation) for supporting through Project-ID: 390740016 (Germany's Excellence Strategy EXC 2075/1, Project PN2-2(II)). TR and HMT thank the Deutsche Forschungsgemeinschaft (DFG, German Research Foundation) for supporting via Project-ID: 465194077 (Priority Programme SPP 2311, Project SimLivA). TR thanks further the Deutsche Forschungsgemeinschaft (DFG, German Research Foundation) for supporting via the following projects: Project-ID: 312860381 (Priority Program SPP 1886, Subproject 12); Project-ID: 436883643 (Research Unit Programme FOR 5151 (QuaLiPerF, Project P7 Liver Lobule); Project-ID: 327154368 (SFB 1313 Project C03 Vertebroplasty); Project-ID: 463296570 (Priority Programme SPP 1158, Antarctica), 504766766 (Project Hybrid MOR), Project-ID: 498601949 (TRR 364, Syntrac, Project B06). TR and HMT are further supported by the Federal Ministry of Education and Research (BMBF, Germany) within ATLAS by grant number 031L0304A and 031L0304C. In addition, TR is supported via the European Union and the German Federal Ministry for Economy and Climate Protection within the framework of the economic stimulus package no. 35c in module b on the basis of a decision by the German Bundestag by the project "DigiTain – Digitalization for Sustainability". RP is supported by the Add-on Fellowship of the Joachim Herz Foundation.